# A warping function-based control chart for detecting distributional changes in damage-sensitive features for structural condition assessment

Zhicheng Chen[1,2], Wenyu Chen[1,2], Xinyi Lei[3*]

[1]School of Mathematics and Statistics, Nanning Normal University, Nanning, 530100, China
[2]Center for Applied Mathematics of Guangxi, Nanning Normal University, Nanning, 530100, China
[3]School of Civil Engineering, Nanyang Institute of Technology, Nanyang, 473004, China

**Abstract**

Data-driven damage detection methods achieve damage identification by analyzing changes in damage-sensitive features (DSFs) derived from structural health monitoring (SHM) data. These methods are more widely applicable as they avoid reliance on complex physical models. The core reason for their effectiveness lies in the fact that damage alters the physical properties of the structure, thereby modifying the data-generation mechanism of its responses. Consequently, the DSF data before and after damage follow different probability distributions. As a result, damage or structural state transition can be manifested as changes in the distribution of DSF data. This enables us to reframe the problem of damage detection as one of identifying these distributional changes. Hence, developing automated tools for detecting such changes is pivotal for automated structural health diagnosis. Control charts are extensively utilized in SHM for DSF change detection, owing to their excellent online detection and early warning capabilities. However, conventional methods are primarily designed to detect mean or variance shifts, making it challenging to identify complex shape changes in distributions. This limitation results in insufficient damage detection sensitivity. Moreover, they typically exhibit poor robustness against data contamination. This paper proposes a novel control chart to address these limitations. It employs the probability density functions (PDFs) of subgrouped DSF data as monitoring objects, with shape deformations characterized by warping functions. Furthermore, a nonparametric control chart is specifically constructed for warping function monitoring in the functional data analysis framework. Key advantages of the new method include the ability to detect both shifts and complex shape deformations in distributions, excellent online detection performance, and robustness against data contamination. Extensive simulation studies demonstrate its superiority over competing approaches. Finally, the method is applied to detecting distributional changes in DSF data for cable condition assessment in a long-span cable-stayed bridge, demonstrating its practical utility in engineering.

*Corresponding author:
Email: xinyileinyist@163.com

## 1. Introduction

Civil engineering structures inevitably accumulate damage throughout their service lives. Such damage may escalate to catastrophic failure, incurring substantial economic losses and social impacts. Consequently, timely damage identification and condition assessment of critical infrastructure are imperative. Structural health monitoring (SHM) has emerged as a pivotal solution to address this challenge [1-3]. By deploying various sensors, SHM systems enable real-time monitoring of structural responses and environmental conditions. The acquired data are then processed to detect changes in material and geometric properties of the structure, facilitating comprehensive structural integrity evaluation. As a fundamental task of SHM, structural damage detection and condition assessment continue to constitute a primary research focus within the SHM community.

In the literature on structural damage detection or condition assessment, most existing methods can be broadly classified into two main categories: model-based methods and data-driven methods [4-7]. While the former demands sophisticated finite element models and substantial computational resources, thereby limiting its practical applicability, the latter does not rely on physical models of structures. Instead, data-driven methods achieve damage detection by analyzing feature changes associated with structural damage in monitoring data, making their implementation more convenient. In practice, meaningful data variations that distinguish damaged from undamaged structures are often obscured in raw monitoring data. Consequently, features sensitive to damage but less affected by environmental or operational changes must be extracted from raw data. Such features are termed damage-sensitive features (DSFs) [1,8], and their pattern changes can indicate structural damage. Therefore, data-driven methods can typically be implemented within the statistical pattern recognition paradigm [1,9,10], with the main process being decomposed into three fundamental steps: data preprocessing, DSF extraction, and feature change detection. The pattern change of DSF data essentially stems from a change in the data's probability distribution. This phenomenon arises because structural damage alters the system's physical properties, thereby modifying the data-generation mechanism of structural responses. Consequently, DSF data exhibit distinct probability distributions before and after damage events. This implies that structural damage or condition changes can be effectively manifested as distributional changes in DSFs. Consequently, detecting distributional changes in DSF data becomes pivotal for data-driven damage detection. Currently, a significant portion of research in SHM has concentrated on data preprocessing and DSF extraction (see Doebling et al. [6] and Sun et al. [7] for comprehensive reviews). However, statistical methods for detecting distributional changes in DSFs remain relatively underexplored.

In SHM applications, the control chart is a pivotal statistical tool for detecting changes in DSF data [11-36]. By employing statistical methods to dynamically analyze continuously collected data, control charts offer significant advantages for real-time variation monitoring and early warning of anomalies. Consequently, they are particularly well-suited for processing streaming data (e.g. sensing data). In addition to control charts, other representative statistical methods employed to discriminate features between undamaged and damaged structures include regression analysis [37-39], change-point detection [40-42], and non-control chart-based outlier detection algorithms [4,43]. Essentially, these methods also relies on analyzing the distributional changes in DSF data to enhance damage detection, but their online detection capabilities are often inferior to those of control charts. For control chart construction, a statistical quantity must be derived from monitoring data to reflect structural performance; this quantity is called the charting statistic [44]. A common practice is to first divide the DSF data into subgroups (referred to as subgrouped data [44]) and then select appropriate statistics (e.g. sample mean, sample variance) from the subgrouped data as the charting statistic. The working principle of control charts is to effectively distinguish between common-cause variation (inherent noise) and special-cause variation (indicative of a change) in the data by establishing statistical control limits derived from the expected behavior of the data under normal conditions. When the charting statistic remains within these control limits, it indicates that only common-cause variation (inherent noise) is present. However, when the charting statistic exceeds the control limits, it indicates the presence of special-cause variation, which signals a meaningful change (such as damage) in the structure, prompting an anomaly alarm to be issued.

Control charts, initially developed for statistical process control, have been widely adapted for SHM applications. Among these, the Shewhart chart is the most commonly used. Two representative types of Shewhart charts are: (a) the X-bar chart for monitoring mean values [11-21] and (b) the S chart for tracking variability [11,12]. These charts are extensively applied to various SHM tasks, including damage detection [11-15], material degradation assessment [16], sensor degradation monitoring [17], anomaly alarms for structural responses [18,19], scour monitoring [20], and welded joints monitoring [21]. Shewhart control charts are simple and easy to use, but they are generally effective only in detecting moderate or large shifts, while being less sensitive to subtle or weak changes. The exponentially weighted moving average (EWMA) control charts have emerged as a widely adopted alternative. These charts demonstrate superior capability in detecting minor data shifts. However, their control limits are typically derived under constant variance assumptions [44], which may not hold true in practical scenarios. Representative studies on structural damage detection using EWMA control charts include Behnia et al. [22], Chaabane et al. [23], Sarwar and Cantero [24], and Wang and Ong [25,26]. In addition to Shewhart and EWMA control charts, another widely used class of control charts is the Hotelling $T^2$ chart. This type of chart is typically employed when multiple variables need to be monitored simultaneously. For these charts, the control limits are derived from either the theoretical or empirical distribution of the $T^2$ statistic. The theoretical approach is based on the assumption of a multivariate normal distribution, whereas the empirical approach requires sufficient historical data collected before a structural change to serve as training samples. In structural damage detection, representative studies employing a Hotelling $T^2$ control chart as the anomaly detection tool include contributions by Magalhães et al. [27], Tondreau and Deraemaeker [28], Diord et al. [29], Comanducci et al. [30], Prakash and Narasimhan [5], Tomé et al. [31], Giglioni et al. [32], García-Macías and Ubertini [33], and Pirrò et al. [34]. Recently, Kullaa [35,36] proposed an extreme value statistics-based control chart that offers advantages in outlier detection. This method involves fitting an extreme value distribution to the block minima or maxima of the data, with control limits derived from the fitted distribution. However, a potential limitation of this approach is that it discards all data points except the maximum or minimum from each block, which may result in underutilization of information.

Control charts have been widely and successfully applied across various SHM tasks. However, conventional control charts exhibit significant limitations when detecting complex distributional changes. Currently, most control charts can only effectively identify changes in mean or variance. For example, X-bar control charts, EWMA control charts, and $T^2$ control charts are primarily designed for mean shift detection, while S control charts specialize in variance shift detection. However, the distributional changes in real SHM data are far more intricate than mere shifts in mean or variance. They may also encompass complex changes in the distribution's shape, including changes in symmetry, variations in tail thickness, increases or decreases in the number of peaks, and shifts in the relative heights between different peaks, among other shape-related characteristics. Unfortunately, conventional control charts face challenges in detecting such general shape changes in SHM data distribution. For instance, the time series generated by the algorithm in Appendix 1 demonstrates no significant changes in either the mean or variance. However, the probability distributions for generating the first and second halves of the data exhibit marked differences in shape. To illustrate this, the dataset is partitioned into 200 equal-length subgroups, and kernel density estimation is applied to estimate the probability density function (PDF) for each subgroup, yielding a PDF sequence denoted as $\{f_i(x)\}_{i=1}^{200}$. The curves of the PDFs are visualized in Fig. 1. Clearly, the shapes of the first half of the PDFs (i.e. $\{f_i(x)\}_{i=1}^{100}$) differ significantly from those of the second half (i.e. $\{f_i(x)\}_{i=101}^{200}$). Under these circumstances, conventional control charts designed for detecting mean or variance shifts become ineffective in identifying meaningful changes in the data. For instance, Fig. 2 presents the X-bar and S control charts constructed for the subgroup means and subgroup standard deviations, respectively. As shown, no data points fall outside the control limits, indicating that no significant changes were detected in the data. In SHM applications, when structural damage causes substantial changes in the shape of the data distribution, while the mean and variance show no significant deviation or their changes are less pronounced than the shape deformation, conventional control charts may have limited capability in detecting such distributional changes. This can reduce their sensitivity in damage detection and even create detection blind spots. Therefore, there is an urgent need to develop new control charts that can address more general distributional changes to overcome the limitations of conventional approaches.

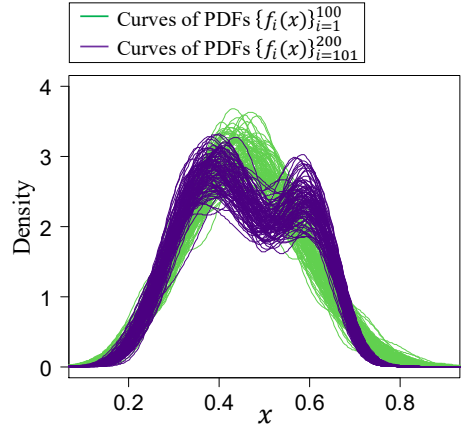

**Fig. 1.** Curves of the PDF-valued samples in the sequence $\{f_i(x)\}_{i=1}^{200}$ estimated from subgrouped data generated by the algorithm in Appendix 1. The first half of the PDFs are represented by light lines, and the second half by dark lines.

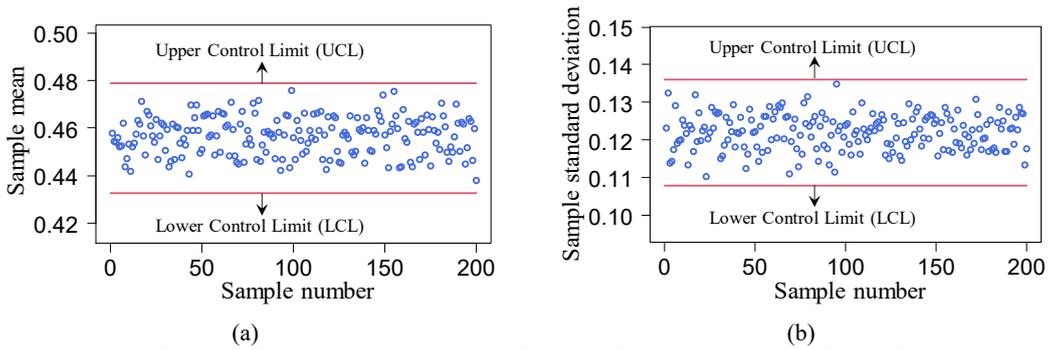

(a)          (b)

**Fig. 2.** Conventional control charts constructed for the subgrouped data: (a) X-bar control chart for subgroup means (hollow circles) and (b) S control chart for subgroup standard deviations (hollow circles). The control limits are indicated by horizontal lines. The sample means and standard deviations are calculated from subgrouped data, obtained by partitioning the raw data generated by the algorithm in Appendix 1 into 200 equal-length subgroups.

On the other hand, real-world SHM systems often operate in complex environments. As a result, the collected monitoring data are inevitably contaminated by various disturbances, leading to the presence of outliers. Consequently, statistical tools designed for SHM applications must demonstrate good robustness. However, most conventional control charts fall short in this regard, resulting in a significantly elevated false alarm rate under outlier interference. Especially the transient anomaly phenomenon shown in Fig. 3, which is characterized by data quickly reverting to normal after a short-term departure from the normal pattern. Such anomalies are typically caused by external interference factors rather than structural damage (data anomalies induced by damage often exhibit persistence and irreversibility). Conventional control charts usually lack effective mechanisms to suppress the influence of such transient anomalies, making them prone to triggering false alarms.

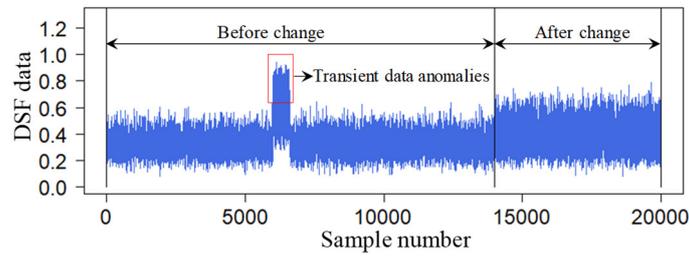

**Fig. 3.** Illustration of transient data anomalies. The rectangle highlights the region where these anomalies occur.

To enhance the detection capability for changes in distribution shape and improve robustness against outliers, this paper proposes a new control chart for real-time monitoring of the distribution of DSF data in SHM. Unlike traditional approaches that utilize the mean or variance of subgrouped data as monitoring variables, this method employs the PDFs of subgrouped data as the primary monitoring object. Building on this foundation, this study develops a control chart specifically for detecting shifts or deformations in PDF curves. A key innovation of this approach lies in the utilization of warping functions to characterize the shape deformation of distributions. A warping function is a special monotonic, smooth, and invertible function that can deform one PDF into another while ensuring the process remains reversible [45-47]. Using this tool, the shape deformation of DSF data's PDFs can be extracted and represented as warping functions. Importantly, both distribution shifts and deformations can be reflected as changes in the warping function. Leveraging the emerging functional data analysis technique, this paper develops a control chart specifically designed to monitor warping functions of probability distributions. This chart allows for simultaneous detection of shifts and shape deformations in the distribution of DSF data. Additionally, this paper uses a rank-based method (a non-parametric statistical analysis approach that relies on data ranking) to construct the control chart, which is less sensitive to outliers. On the other hand, unlike traditional methods that perform statistical analysis on the mean or variance of subgrouped data, this paper directly analyzes the probability distributions (i.e. the PDFs and their warping functions) of the subgrouped data. This strategy can also supresss the influence of outliers. Furthermore, the rank-based control chart for warping function monitoring in this paper can flexibly control the sensitivity to short-term transient anomalies by adjusting the smoothing parameter. This capability ensures robust detection performance even when data contamination occurs.

## 2. Data subgrouping and distribution summary

The core purpose of this paper is to develop a novel control chart for detecting distributional changes in DSF data. This detection process corresponds to the third step of the statistical pattern recognition paradigm mentioned in the introduction, namely feature change detection. It should be noted that the three main steps of this paradigm (i.e. data preprocessing, DFS extraction, and feature change detection) can be executed sequentially. The data preprocessing in the first step is also referred to as data normalization in some literature. Its primary purpose is to eliminate confounding effects from raw monitoring data that are induced by environmental and operational factors. Given that a wealth of well-established methods already exists for the first two steps, this paper does not propose new data processing approaches for them. Instead, it is assumed that the first two steps have been completed using existing methods, and the corresponding DSF data required for control chart construction have been obtained in advance.

let $\boldsymbol{X} = \{X_1, X_2, \cdots\}$ denote the time series of extracted DSF data. These data are subsequently partitioned into subgroups, known as subgrouped data [44]. When the data is segmented into equal-length subgroups, the $j$-th subgroup can be expressed as $\boldsymbol{X}_j^G = \{X_{(j-1)m+1}, \cdots, X_{jm}\}$ (for $j = 1, 2, \cdots$), where $m$ represents the subgroup size (i.e. the number of data points per subgroup). In practical applications, other segmentation methods (not limited to equal-length segmentation) can also be employed based on specific requirements, such as grouping data by hourly or daily intervals. It is worth noting that dividing the raw data into subgroups before constructing control charts is a commonly adopted approach in statistical process control (SPC) [44]. In traditional control chart-based monitoring approaches for subgrouped data, statistics derived from these subgroups (e.g. the sample mean, sample variance, or other relevant parameter estimates) are routinely employed as monitoring objects. However, relying solely on a single parameter or a few parameters to describe subgrouped data, traditional approaches may lead to significant information loss, particularly in capturing the distributional shape characteristics of the data. To address this limitation, this study employs an alternative strategy. The PDF of each subgrouped data is first estimated using the kernel density estimation technique, yielding a sequence of PDFs denoted as $\mathcal{F} = \{f_1(x), \cdots, f_j(x), \cdots\}$, where $f_j(x)$ is the PDF estimated from the $j$-th subgrouped data. Subsequently, the PDFs themselves are utilized as the monitoring objects. This strategy, referred to as distribution summary in this study (see Fig. 4 for a schematic illustration), not only preserves the distribution information of the data but also effectively suppresses the influence of outliers in the original DSF data. Outliers that deviate significantly from the data center can markedly influence the computation of the sample mean or variance. However, when the number of outliers is small, their impact on the overall shape of the data's probability distribution is typically limited. Yet, if a large number of outliers occur consecutively (e.g. the transient anomaly phenomenon depicted in Fig. 3), they can be condensed into a few outlying PDF samples after undergoing the distribution summary process (see Fig. 5).

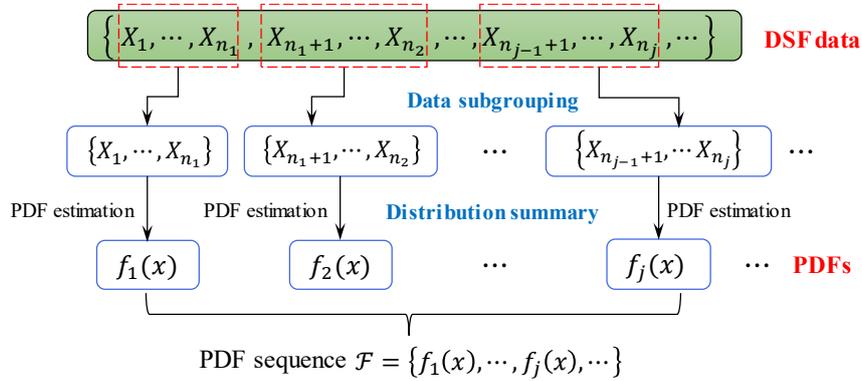

**Fig. 4.** Illustration of data subgrouping and distribution summary.

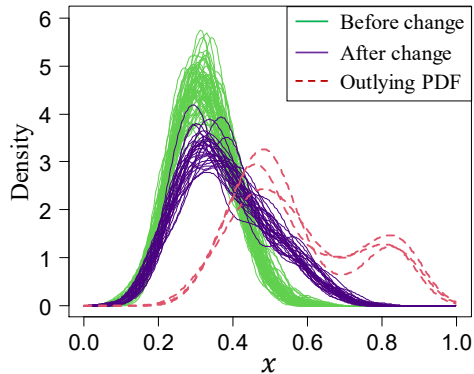

**Fig. 5.** PDFs resulting from the data in Fig. 3 after distribution summary. The time series in Fig. 3 is partitioned into 100 equal-length subgroups; the three outlying PDFs (marked by bold dashed lines) correspond to the transient anomalous data.

In the monitoring process, if no data subgrouping is performed and each data point of the DSF data is individually inspected using a control chart, such data is termed individual data in the SPC literature, and the chart is called an individual control chart [44]. Generally, this approach is suitable for scenarios with long data acquisition cycles, high measurement costs, or have difficulties in data division [44]. However, the DSF data extracted from SHM data typically contains a large number of sample points. The individual data inspection approach is generally not advisable for two primary reasons: first, the large sample size increases computational complexity, and second, monitoring individual values is more susceptible to noise and outliers, significantly elevating the risk of false alarms. Additionally, individual data does not contain information about the distribution shape of the data. Therefore, this paper does not adopt this method but instead employs the aforementioned data subgrouping and distribution summary strategy.

After performing the distribution summary, the distributional detection problem of the DSF data can be reformulated as detecting changes in the PDF sequence $\mathcal{F} = \{f_1(x), \cdots, f_j(x), \cdots\}$. The control chart to be developed in the next section is specifically designed for this purpose: to distinguish between common-cause variation (inherent noise) and special-cause variation (indicative of a change) in the PDF-valued data. The detected distributional change can signify a potential damage that has occurred in the structure.

The PDFs in $\mathcal{F} = \{f_1(x), \cdots, f_j(x), \cdots\}$ are assumed to be take nonzero values within a common finite interval [a, b], i.e. $f_j(x) > 0$ for $x \in [a, b]$ and $f_j(x) = 0$ for $x \notin [a, b]$. This assumption is generally satisfied in SHM applications because the extracted DSF data has finite values. The interval [a, b] is referred to as the support of the PDF. For a detailed discussion on estimating the support of the PDF using the investigated data, see [48]. Additionally, an invertible scale transformation can be applied to convert a PDF supported on any finite interval [a, b] into a PDF supported on [0, 1] (see Appendix 4 of Chen et al. [48] for details). Since this transformation is invertible, the PDF supported on [0, 1] can be scaled back to its original support [a, b]. Therefore, without loss of generality, the supports of all PDFs are assumed to be [0, 1] when developing the related control chart for detecting changes in the PDF-valued sequence; this assumption is referred to as the 'unit support assumption' in this study.

## 3. Warping function-based control chart

After implementing the data subgrouping and distribution summary described in the previous section, the distributional information of the subgrouped DSF data is represented by the PDFs. The data to be investigated are the PDF-valued data in the PDF sequence $\mathcal{F} = \{f_1(x), \cdots, f_j(x), \cdots\}$.

Such data are referred to as distributional data throughout the rest of this study. In statistics, distributional data refers to data composed of functions related to probability distributions [49], such as PDFs and cumulative distribution functions (CDFs).

The PDF sequence $\mathcal{F} = \{f_1(x), \cdots, f_j(x), \cdots\}$ of the DSF data can be regarded as a realization of a PDF-valued stochastic process. Consequently, to monitor changes in the underlying PDF-valued stochastic process, one must select appropriate process characteristics for monitoring. These characteristics should be sensitive to both positional shifts and shape deformations in the distributional data. To achieve this objective, this study proposes monitoring the warping functions derived from the distributional data rather than directly monitoring the PDFs.

### *3.1. Warping function computation*

In mathematics, a warping function is defined as an invertible, smooth mapping from the interval [0,1] to itself, satisfying the boundary conditions $\gamma(0) = 0$ and $\gamma(1) = 1$. The set of all such warping functions is denoted as [45-47]

$$\boldsymbol{\Gamma} = \{\gamma: [0,1] \to [0,1] \mid \gamma(0) = 0, \gamma(1) = 1, \gamma \text{ is invertible, and both } \gamma \text{ and } \gamma^{-1} \text{are smooth}\} \quad (1)$$

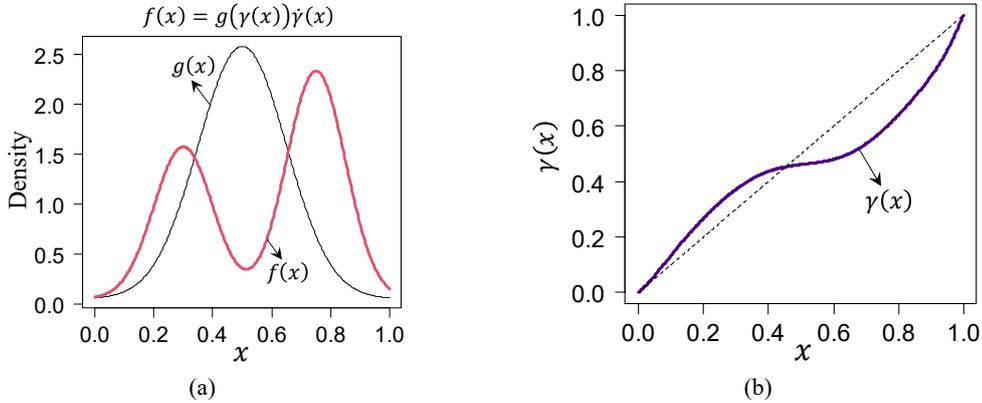

(a)          (b)

**Fig. 6.** Warping transformation for converting a PDF $g(x)$ into another PDF $f(x)$. (a) PDFs $g(x)$ (before warping) and $f(x)$ (after warping); (b) Warping function $\gamma(x)$ satisfying $f(x) = g(\gamma(x))\dot{\gamma}(x)$.

Based on the warping transformation theory of probability distributions, for any two arbitrary smooth PDFs $f(x)$ and $g(x)$ supported on [0,1], there exists a unique warping function $\gamma(x) \in \boldsymbol{\Gamma}$ that transforms the PDF $g(x)$ into the PDF $f(x)$ as follows [46]:

$$f(x) = g(\gamma(x))\dot{\gamma}(x) \quad (2)$$

where $\dot{\gamma}(x)$ denotes the derivative of $\gamma(x)$ with respect to $x$. For a visual demonstration, see Fig. 6. The warping function can be viewed as an operation that deforms the shape of the PDF $g(x)$ into the PDF $f(x)$. The transformation in Eq. (2) is termed the warping transformation of distributions. It is worth noting that this transformation is invertible, meaning that the PDF $f(x)$ can be transformed back into $g(x)$ using the inverse warping function $\gamma^{-1}(x)$.

The warping transformation for distributions defined in Eq. (2) can be understood more intuitively through the lens of CDFs. Let $F(x)$ and $G(x)$ denote corresponding CDFs of the PDFs $f(x)$ and $g(x)$, respectively. The warping function $\gamma(x)$, which morphs the PDF $g(x)$ into $f(x)$ through shape deformation, satisfies the equality

$$F(x) = G(\gamma(x)) \tag{3}$$

Taking the derivative of both sides of this equation with respect to $x$ reveals the underlying PDF relationship specified in Eq. (2). Fundamentally, the warping function $\gamma(x)$ transforms the CDF $G(x)$ by warping its domain along the x-axis, thereby producing $F(x)$. In fact, $\gamma(x)$ represents the phase component of $G(x)$ with respect to $F(x)$ [45]. Consequently, both the position shift and shape deformation of distributions can be captured by the warping function. In other words, the warping function is sensitive not only to the shape deformation of the distribution but also to its position shift. This can be verified by comparing the corresponding warping functions to the identity warping function. The identity warping function is defined as $\gamma_e(x) = x$, $x \in [0,1]$. If a PDF is warped by the identity warping function, it remains unchanged (this can be easily verified by substituting $\gamma_e(x)$ into Eq. (2)). Fig. 7 presents the warping function for the position shift case, with the identity warping function $\gamma_e(x)$ superimposed for comparison. Since the warping function in the position shift case also significantly deviates from the identity warping function $\gamma_e(x)$ (corresponding to the case of no distributional change), this indicates that the position shift of a distribution can also be manifested as a change in the corresponding warping function. Consequently, the distributional change detection problem of the DSF data can be further shifted to a problem of detecting changes in the warping functions.

The domain warping principle (Eq.(3)) enables direct computation of the warping function as follows [45]:

$$\gamma(x) = G^{-1}(F(x)) \tag{4}$$

where $G^{-1}$ is the inverse function of the CDF $G(x)$, also known as the quantile function.

-12-

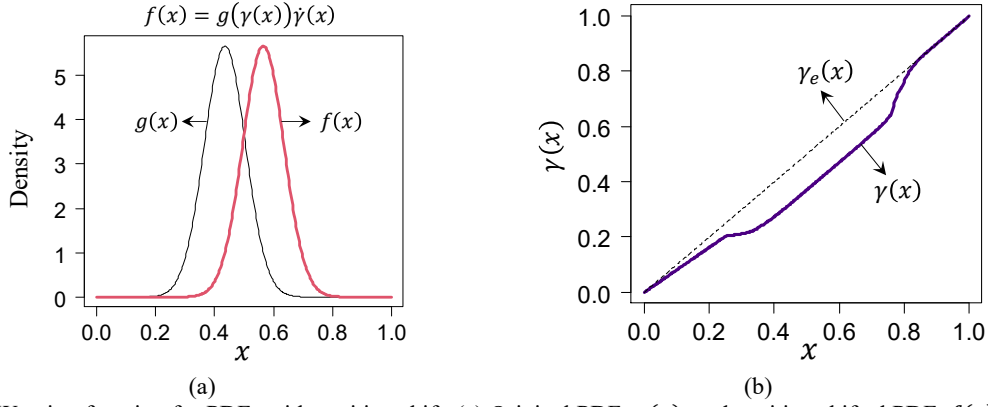

**Fig. 7.** Warping function for PDFs with position shift. (a) Original PDF $g(x)$ and position shifted PDF $f(x)$; (b) Warping function $\gamma(x)$ satisfying $f(x) = g(\gamma(x))\dot{\gamma}(x)$.

The initial $N_0$ probability distributions within the distributional sequence are assumed to be generated from the pre-change regime. These samples are referred to as training samples, which can be used to construct a reference distribution for extracting warping functions for process monitoring. Specifically, let $F_1(x), F_2(x), \cdots, F_{N_0}(x)$ denote the CDFs of the training distributions, with their corresponding quantile functions represented as $F_1^{-1}(x), F_2^{-1}(x), \cdots, F_{N_0}^{-1}(x)$. Then, the reference function is determined as the CDF $F_*(x)$ associated with the sample mean quantile function $F_*^{-1}(x) = \frac{1}{N_0}\sum_{i=1}^{N_0} F_i^{-1}(x)$. Both historical and future CDFs are warped to this reference CDF $F_*(x)$. Using Eq.(4), the corresponding warping functions can be computed as

$$\gamma_i(x) = F_i^{-1}(F_*(x)), i = 1, 2, \cdots, N_0, N_0 + 1, \cdots \tag{5}$$

It is worth noting that for PDFs with near-zero values in a subinterval $I = (a, b) \subset [0,1]$, numerical instabilities may arise when computing warping functions using Eq. (5). For more details please refer to the supplementary materials of Lei et al. [50]. A simple yet effective solution is to preprocess the PDFs $\{f_i(x)\}_{i\geq 1}$ into the following form before computing their corresponding CDFs [50]:

$$f_i(x) = (1 - \alpha_{mix})f_i(x) + \alpha_{mix}\chi_{[0,1]}(x), i = 1, 2, \cdots, N_0, N_0 + 1, \cdots \tag{6}$$

where $\chi_{[0,1]}(x)$ denotes the PDF of the uniform distribution on $[0,1]$, defined as $\chi_{[0,1]}(x) = 1$ for $x \in [0,1]$ and 0 otherwise, $\alpha_{mix} \in [0.1, 0.3]$ is the mixing coefficient. In this study, unless otherwise stated, all PDFs are preprocessed using Eq. (6) with $\alpha_{mix} = 0.1$ prior to warping function computation.

With the above processing, a warping function-valued sequence $\mathcal{Z} = \{\gamma_1(x), \cdots, \gamma_j(x), \cdots\}$ associated with the distributional sequence $\mathcal{F} = \{f_1(x), \cdots, f_j(x), \cdots\}$ can be obtained. The remaining task is to develop a control chart specifically for monitoring the process of warping functions. The warping functions $\{\gamma_i(x)\}_{i\geq 1}$ are treated as random samples of a functional process, and thus they naturally fall into the scope of functional data. Consequently, the control chart can be developed within the framework of functional data analysis (FDA). FDA is a modern branch of statistics dedicated to handling infinite-dimensional data such as functions, curves, or surfaces [51,52]. It is worth noting that the warping function space (i.e. the set $\Gamma$ defined in Eq.(1)) is a nonlinear space, which is not closed under linear operations such as addition and scalar multiplication [47]. This mathematical property creates substantial analytical and computational challenges for subsequent data processing and control chart construction. To address this challenge, this study adopts a transformation-based approach. Specifically, the warping functions are first transformed into a suitable linear space using appropriate functional data transformation tools. Subsequently, dimension reduction is applied to the transformed data based on functional principal component analysis (FPCA), and monitoring statistics are constructed based on the dimensionality reduction results. Finally, a rank-based nonparametric control chart is employed for process monitoring. Combining the warping function computation process described in this subsection, the main steps of the proposed warping function-based control chart method for monitoring distributional changes are summarized in the flowchart in Fig. 8, with the remaining three steps detailed in the following subsections.

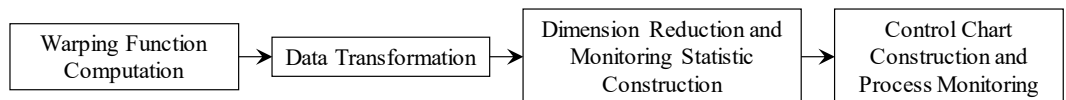

**Fig. 8.** The overall process of constructing the warping function-based control chart.

## *3.2. Data transformation*

The data transformation for warping functions consists of two stages: first, converting warping functions to square-root slope functions (SRSFs); second, transforming these SRSFs into a linear space. For any given warping function $\gamma(x) \in \Gamma$, the corresponding SRSF is defined as $q(x) = \sqrt{\frac{d}{dx}\gamma(x)}, x \in [0,1]$ [45,46]. Since warping functions are absolutely continuous on [0, 1], their SRSFs are square-integrable. As a result, the space of SRSFs forms a subspace of $L^2([0,1])$ (i.e. the set of all square-integrable functions defined on [0,1] ). Under the $L^2$-norm (i.e.

$\|u\| = \left(\int_0^1 |u(t)|^2 dt\right)^{1/2}$), SRSFs satisfy the unit norm condition $\|q\| = \left(\int_0^1 |q(t)|^2 dt\right)^{1/2} = \gamma(x)|_0^1 = 1$. This condition guarantees that SRSFs always lie within the positive orthant of the Hilbert unit sphere (HUS) [45,46] defined as follows:

$$S_\infty = \{u \in L^2([0,1]) \mid d_{L^2}(u, \mathbf{0}) = \|u\| = 1\} \tag{7}$$

where $d_{L^2}$ denotes the $L^2$-distance, and $\mathbf{0}$ represents the zero element of the $L^2([0,1])$ space. By representing warping functions as SRSFs, the complicated geometry of the space $\Gamma$ is reduced to a unit sphere $S_\infty$. Notably, $S_\infty$ has a geometrically favorable property: its tangent space at any point is a linear space [45]. The tangent space of the HUS $S_\infty$ is a generalization of the tangent plane to a sphere in 3D Euclidean space. Existing transformations are available in statistical literature for transforming SRSFs onto a prescribed tangent space. Let $q_e(x) = 1$, $x \in [0,1]$ denote the SRSF of the identify warping function $\gamma_e(x) = x$, $x \in [0,1]$, the tangent space of $S_\infty$ at $q_e(x)$ (denoted as $T_e(S_\infty)$) is selected as the tangent space for data transformation. Combining the SRSF representation, the transformation for converting warping functions from $\Gamma$ to the tangent space $T_e(S_\infty)$ is given by [45,46]

$$v(x) = \frac{\theta}{\sin(\theta)}(q(x) - q_e(x) \cdot \cos(\theta)) \text{ with } q(x) = \sqrt{\frac{d}{dx}\gamma(x)}, \quad \forall \gamma \in \Gamma \tag{8}$$

where $\theta = \cos^{-1}(\int q_e(t) q(t) dt)$. It is worth noting that this transformation is invertible, meaning $v(x)$ can be transformed back into the warping function $\gamma(x)$ without information loss. Since the method in this paper does not require the use of the inverse transformation, the specific form of the inverse transformation is not provided here. Interested readers may refer to [46].

Using Eq. (8), the extracted warping functions (Eq. (5)) are transformed into the tangent space $T_e(S_\infty)$, yielding the results denoted as $\{v_i(x)\}_{i=1}^n$, which are referred to as $T_e(S_\infty)$-representations of the warping functions. Fig. 9 visualizes both the warping functions and their $T_e(S_\infty)$-representations for the distributional dataset shown in Fig. 1. The results in Fig. 9 demonstrate that after transforming the warping functions into $T_e(S_\infty)$-representations, the distinction between data before and after the change becomes more pronounced.

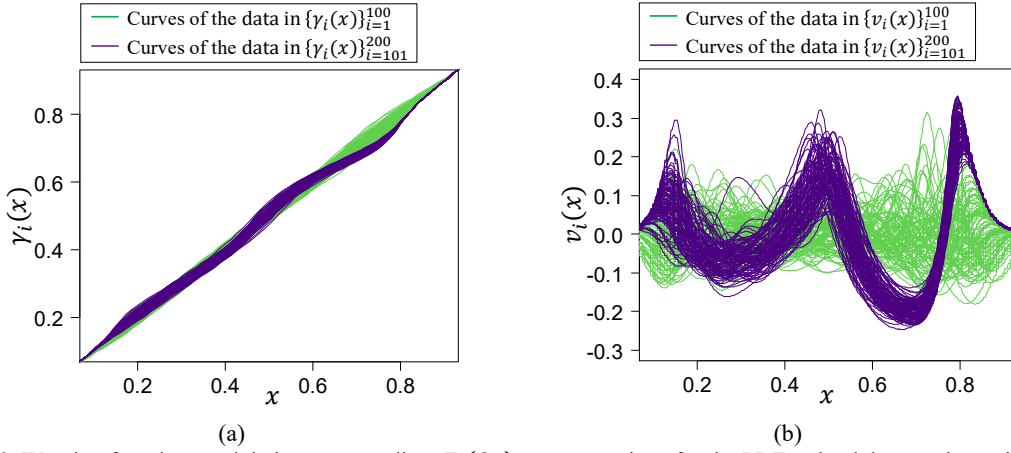

(a)                                       (b)

**Fig. 9.** Warping functions and their corresponding $T_e(S_\infty)$-representations for the PDF-valued dataset shown in Fig. 1. (a) Warping functions and (b) $T_e(S_\infty)$-representations of warping functions. Light curves correspond to data before pattern change (i.e. $\{\gamma_i(x)\}_{i=1}^{100}$ or $\{v_i(x)\}_{i=1}^{100}$), while dark curves correspond to data after pattern change (i.e. $\{\gamma_i(x)\}_{i=101}^{200}$ or $\{v_i(x)\}_{i=101}^{200}$).

### 3.3. Dimension reduction and monitoring statistic construction

The tangent space $T_e(S_\infty)$ is an infinite-dimensional linear space. The intrinsic infinite-dimensionality of the $T_e(S_\infty)$-representations of warping functions poses significant challenges for subsequent monitoring statistics construction. To address this, this study employs FPCA [51,52] to perform dimensionality reduction on the $T_e(S_\infty)$-representations of warping functions, upon which the monitoring statistics are derived from the FPCA results.

The primary aim of FPCA is to establish an empirical orthonormal basis that offers the most concise representation of functional data. The $T_e(S_\infty)$-representations of the training distributional data (i.e. the first $N_0$ probability distributions assumed to precede structural changes) are used to construct the FPC model. Let $\{v_i(x)\}_{i=1}^{N_0}$ denote the $T_e(S_\infty)$-representations corresponding to the training data, with the sample mean function defined as $\bar{v}_{1:N_0}(x) = \frac{1}{N_0}\sum_{i=1}^{N_0} v_i(x)$. In FPCA, these representations are treated as realizations of a functional random variable $v$. The empirical covariance function of $v$ is then formulated as [51,52]

$$c(s,t) = \frac{1}{N_0}\sum_{i=1}^{N_0}\left\{\left(v_i(s) - \bar{v}_{1:N_0}(s)\right)\left(v_i(t) - \bar{v}_{1:N_0}(t)\right)\right\} \tag{9}$$

The covariance operator $\mathbb{K}$ is defined through the integral transform $(\mathbb{K}v)(s) = \int c(s,t)v(t)dt$. Let $\{\varphi_l(x)\}$ represent the eigenfunctions of $\mathbb{K}$ corresponding to non-ascending eigenvalues $\{\rho_l\}$, satisfying [51,52]

$$(\mathbb{K}\varphi_l)(s) = \int c(s,t)\varphi_l(t)dt = \rho_l\varphi_l \text{ and } \int \varphi_l^2(t)dt = 1, \ l = 1,2,\cdots \quad (10)$$

For computational details of solving this eigenvalue problem, see Ramsay and Silverman [51]. According to FPCA theory [52], $\{\varphi_l(x)\}$ forms an orthonormal basis that enables representation of the $T_e(S_\infty)$-representations $\{v_i(x)\}_{i=1}^{N_0}$ as follows:

$$v_i(x) = \bar{v}_{1:N_0}(x) + \sum_{l=1}^{\infty} \beta_{il}\,\varphi_l(x), \qquad i = 1,\cdots,N_0 \quad (11)$$

where $\{\varphi_l(x)\}$ are termed FPCs, and $\beta_{il} = \int \left(v_i(\tau) - \bar{v}_{1:N_0}(\tau)\right)\varphi_l(\tau)d\tau$ represents the FPC score of $v_i(x)$ on $\varphi_l(x)$. The expression in Eq.(11) is referred to as the Karhunen–Loève (KL) expansion of $v_i(x)$. By truncating the KL expansion to retain only the first $L$ FPCs with dominant contributions, the infinite-dimensional functional data $v_i(x)$ can be reduced to the following finite-dimensional form:

$$v_i^L(x) = \bar{v}_{1:N_0}(x) + \sum_{l=1}^{L} \beta_{il}\,\varphi_l(x), \qquad i = 1,\cdots,N_0 \quad (12)$$

where $L$ denotes the number of retained PFCs. In practice, $L$ is chosen such that the retained PFCs $\varphi_1,\cdots,\varphi_L$ capture at least a specified proportion $\theta$ of the total variability. Formally,

$$L = \min\left\{k \in \mathbb{N}: \frac{\sum_{l=1}^{k}\rho_l}{\sum_{l=1}^{\infty}\rho_l} \geq \theta\right\} \quad (13)$$

where $\mathbb{N}$ is the set of positive integers, and $\{\rho_l\}$ are the eigenvalues given in Eq. (10). Unless otherwise specified, $\theta$ is set to 99% in this study.

After constructing the FPC model, the $T_e(S_\infty)$-representations of all distributional data in the process (for $i \geq N_0$, including new data from future monitoring) are projected onto this model to compute monitoring statistics. Let $\{v_i(x)\}_{i \geq N_0}$ denote the $T_e(S_\infty)$-representations for $i \geq N_0$. For distributional change monitoring, two complementary statistical measures are employed in this study. The first monitoring statistic is Hotelling's $T^2$ [53], which assesses changes within the principal subspace. This statistic is computed as

$$T_i^2 = \sum_{l=1}^{L} \frac{\beta_{il}^2}{\rho_l}, \qquad i = N_0 + 1, N_0 + 2,\cdots \quad (14)$$

where $\beta_{il} = \int \left(v_i(\tau) - \bar{v}_{1:N_0}(\tau)\right)\varphi_l(\tau)d\tau$ denotes the $l$-th FPC score of $v_i(x)$, and $\{\rho_l\}_{l=1}^{L}$ represents the first $L$ largest eigenvalues from Eq. (10). This metric is specifically designed to capture changes occurring within the principal subspace $\mathcal{W} = \text{span}(\varphi_1, \cdots, \varphi_L)$ of the functional process $\{v_i(x)\}_{i \geq N_0}$. To capture changes occurring outside this principal subspace, the following squared prediction error (SPE) statistics [53] are further constructed:

$$SPE_i = \left\|v_i - \hat{v}_i^L\right\|_{L_2}^2 = \int \left(v_i(\tau) - \hat{v}_i^L(\tau)\right)^2 d\tau, \qquad i = N_0 + 1, N_0 + 2, \cdots \tag{15}$$

where $\hat{v}_i^L(\tau) = \bar{v}_{1:N_0}(x) + \sum_{l=1}^{L} \beta_{il}\, \varphi_l(x)$ represents the reconstruction of $v_i(x)$ based on the $L$ retained FPCs. This complementary metric can capture changes in the orthogonal complement space $\mathcal{W}^\perp$, providing comprehensive coverage of distributional variation across both principal and residual components.

For simplicity of notation, the two statistical quantity sequences $\{T_i^2\}_{i \geq N_0}$ (Eq.(14) and $\{SPE_i\}_{i \geq N_0}$ (Eq.(15) are renumbered starting from $k = 1$ for elements with $i \geq N_0 + 1$. The renumbered sequences are expressed as

$$\tilde{T}_k^2 = T_{N_0+k}^2, \qquad k = 1, 2, \cdots \tag{16a}$$

$$\widetilde{SPE}_k = SPE_{N_0+k}, \qquad k = 1, 2, \cdots \tag{16b}$$

### 3.4. Control chart construction and process monitoring

This subsection constructs the control charts for the Hotelling's $T^2$ statistic process (i.e. $\{\tilde{T}_k^2\}_{k \geq 1}$) and the SPE statistic process (i.e. $\{\widetilde{SPE}_k\}_{k \geq 1}$).

The simplest and most straightforward approach is to use $\tilde{T}_k^2$ (or $\widetilde{SPE}_k$) itself as the charting statistic for constructing the control chart. Since the distribution of $\tilde{T}_k^2$ (or $\widetilde{SPE}_k$) is unknown, a segment of pre-change data is typically required to estimate the control limits of the chart. This data segment is referred to as the tuning data. Assuming $\tilde{T}_k^2$ (or $\widetilde{SPE}_k$) for $k = 1, \cdots, m$ serves as the tuning data, the upper control limit can be set as the sample $1 - \alpha$ quantile of the tuning data, where $\alpha$ represents the overall Type I error probability, similar to that in [53]. This strategy is referred to as the direct charting method in this study. Despite its simplicity, the direct charting method has two limitations: (1) high false alarm rates due to sensitivity to outliers, and (2) the requirement for lengthy tuning data to obtain reliable control limit estimates. To address these challenges, this study employs a rank-based methodology: (1) treating $\tilde{T}_k^2$ or $\widetilde{SPE}_k$ as primary features instead of charting statistics; (2) constructing charting statistics from their ranks; and (3) applying the nonparametric control chart proposed by Zhou et al.[54] for process monitoring. This approach offers three significant advantages: it enables real-time online monitoring, enhances robustness against outliers, and eliminates the need for lengthy tuning data. Notably, the rank-based nonparametric control chart in Zhou et al. [54] is originally designed for mean shift detection. However, by employing the warping function modeling and transformation framework presented in this study, both shifts and shape deformations in the PDF curves of DSF data can be manifested as mean changes in warping functions, enabling unified detection through mean change analysis. This is a key innovation that distinguishes this work from existing approaches.

In the following, we will use $\{\tilde{T}_k^2\}_{k\geq 1}$ as an example to describe the computation procedure for control chart construction. The control chart for $\{\widetilde{SPE}_k\}_{k\geq 1}$ can be established similarly. The first $m$ data points in the sequence $\{\tilde{T}_k^2\}_{k\geq 1}$ are assumed to be historical data, which serve as tuning data. Unlike the direct charting approach, the length of the tuning data can be significantly shorter. Therefore, $m$ can be set to a much smaller value, with its default value typically set to $m = 30$. The data in $\{\tilde{T}_k^2\}_{k\geq 1}$ for $k > m$ is assumed to be future data. Suppose that, up to the present, a total of $n$ future data points have been collected. Combined with the $m$ historical data points, the complete dataset available at the current time consists of $\tilde{T}_1^2, \cdots, \tilde{T}_m^2, \tilde{T}_{m+1}^2, \cdots, \tilde{T}_{m+n}^2$. For any $1 \leq k \leq m + n$, the rank of $\tilde{T}_k^2$ is defined as the number of data points in $\{\tilde{T}_k^2\}_{k=1}^{m+n}$ that are not larger than $\tilde{T}_k^2$ [55]. That is,

$$R_k(\tilde{T}^2) = \sum_{i=1}^{m+n} I\{\tilde{T}_i^2 \leq \tilde{T}_k^2\}, \ k = 1, 2, \cdots, m + n \tag{17}$$

where $I\{\tilde{T}_i^2 \leq \tilde{T}_k^2\}$ is the indicator function defined as follows:

$$I\{\tilde{T}_i^2 \leq \tilde{T}_k^2\} = \begin{cases} 1, & \tilde{T}_i^2 \leq \tilde{T}_k^2 \\ 0, & \tilde{T}_i^2 > \tilde{T}_k^2 \end{cases} \tag{18}$$

The nonparametric control chart for monitoring the process of $\{\tilde{T}_k^2\}_{k\geq 1}$ is constructed based on the Mann-Whitney statistic. To compute this statistic, for any $1 \leq j < m + n$, the Wilcoxon rank-sum statistics [55] are first calculated using the following formula:

$$W_{j,m+n} = \sum_{k=1}^{j} R_k(\tilde{T}^2), 1 \leq j < m + n \tag{19}$$

where $W_{j,m+n}$ represents the Wilcoxon rank-sum statistic at position $j$. Based on the relationship between the Wilcoxon rank-sum statistic and the Mann-Whitney statistic, the corresponding Mann-Whitney statistic can be derived as $MW_{j,m+n} = W_{j,m+n} - j(j+1)/2$ [55]. Subsequently, the following standardized Mann-Whitney (SMW) statistic is computed for process monitoring [54]:

$$SMW_{j,m+n} = \frac{MW_{j,m+n} - E_0(MW_{j,m+n})}{\sqrt{Var_0(MW_{j,m+n})}}, \quad j = 1, 2, \cdots, (m+n) - 1 \tag{20}$$

where $E_0(MW_{j,m+n})$ and $Var_0(MW_{j,m+n})$ denote the expectation and variance of $MW_{j,m+n}$ under the in-control condition, respectively. The analytical expressions for $E_0(MW_{j,m+n})$ and $Var_0(MW_{j,m+n})$ are given as follows [54]:

$$E_0(MW_{j,m+n}) = \frac{j((m+n) - j)}{2} \tag{21a}$$

$$Var_0(MW_{j,m+n}) = C_{tie} \frac{j((m+n) - j)((m+n) + 1)}{12} \tag{21b}$$

where $C_{tie}$ is the tie correction factor for variance. Suppose among the $m+n$ data points $\tilde{T}_1^2, \cdots, \tilde{T}_m^2, \tilde{T}_{m+1}^2, \cdots, \tilde{T}_{m+n}^2$, there are $r$ distinct values, with the $i$th value appearing $\omega_i$ times. Then, $C_{tie}$ can be computed as [54]

$$C_{tie} = 1 - \frac{1}{(n+m)((n+m)^2 - 1)} \sum_{i=1}^{r} \omega_i(\omega_i^2 - 1) \tag{22}$$

Using the computed SMW statistics for the process $\{\tilde{T}_k^2\}_{k \geq 1}$, the rank-based nonparametric control chart in [54] can be applied to conduct online monitoring. In the monitoring procedure, upon the $n$th future sample is collected, compute the weighted moving average (WMA) statistics using the following recursive formula:

$$Y_j(m,n) = \lambda \cdot SMW_{j,m+n} + (1-\lambda) \cdot Y_{j-1}(m,n),$$
$$for\ j = m - m_0, m - m_0 + 1, \cdots, m + n - 1 \tag{23}$$

where $m_0 \in [4,10]$ is an integer selected to reduce instability in the statistics $Y_j(m,n)$ under small future sample sizes ($n$), and $\lambda$ is the smoothing parameter satisfying $0 < \lambda < 1$, $SMW_{j,m+n}$ is the SMW statistic computed using Eq. (20). The initial value of $Y_j(m,n)$ (when $j = n + m_0 - 1$) is set to 0. Then the local maxima of the absolute WMA statistics defined in the following form is used as the charting statistic [54]:

$$Y_{max}(m,n) = \max_{m-m_0 \leq j < m+n} |Y_j(m,n)|, \quad n = 1, 2, \cdots \tag{24}$$

This charting statistic is referred to as the Y-max statistic throughout the rest of this study. The control chart is achieved by plotting values of this charting statistic against the sample index. If the value of $Y_{max}(m,n)$ exceeds the prescribed control limit, an anomaly alarm is triggered, otherwise, the system continues to monitor the $(n+1)$st future sample. In practical applications, the control limit, denoted as $h_{m,n}$, can be obtained by consulting Table 1 in [54] based on the user-specified in-control average run-length (IC ARL) of the chart.

When an anomaly alarm is triggered during the inspection of the $n$th future sample in $\{\tilde{T}_k^2\}_{k \geq 1}$, it indicates a change has occurred in the data. The position of this change can be estimated as [54]

$$\hat{\tau}^* = \underset{m \leq t < m+n}{\operatorname{argmax}} |SMW_{t,m+n}| \tag{25}$$

In statistics, the position of the change in the process is referred to as the change point. Thus, $\hat{\tau}^*$ serves as the estimate of the change point. Note that the sequence $\{\tilde{T}_k^2\}_{k \geq 1}$ obtained using Eq.(16a) excludes the $N_0$ samples associated with the training data. Therefore, if one wishes to locate the change position in the original data sequence, $\hat{\tau}^*$ must be corrected by adding $N_0$, i.e.

$$\hat{\tau} = \hat{\tau}^* + N_0 \tag{26}$$

Since the change in the Hotelling's $T^2$ statistic process $\{\tilde{T}_k^2\}_{k \geq 1}$ indicates a change in the corresponding distributional sequence of the DSF data, the estimated change point can be treated as the change point of the distributional sequence.

Now, the introduction to the control chart construction and change-point estimation for the Hotelling's $T^2$ statistic process (i.e. $\{\tilde{T}_k^2\}_{k \geq 1}$) has been completed. The control chart and change-point estimation for the SPE statistic process (i.e. $\{\widetilde{SPE}_k\}_{k \geq 1}$) can be implemented similarly, and the technical details are omitted here for brevity. These two proposed control charts for detecting and monitoring distributional changes in DSF data are constructed using warping functions and rank-based nonparametric statistical methods. The first chart employs Hotelling's $T^2$ statistic to capture changes within the principal subspace of the $T_e(S_\infty)$- representations of warping functions, while the second chart uses the SPE statistic to detect changes occurring outside the principal subspace. For convenience, throughout the rest of this study, the control charts associated with Hotelling's $T^2$ statistic and the SPE statistic are referred to as the Warp-RANK-$T^2$ and Warp-RANK-SPE charts, respectively. Since data changes can occur both within and outside the principal subspace, these two control charts should be used in combination. If either chart triggers an anomaly alarm, it indicates that the DSF data may have undergone a distributional change.

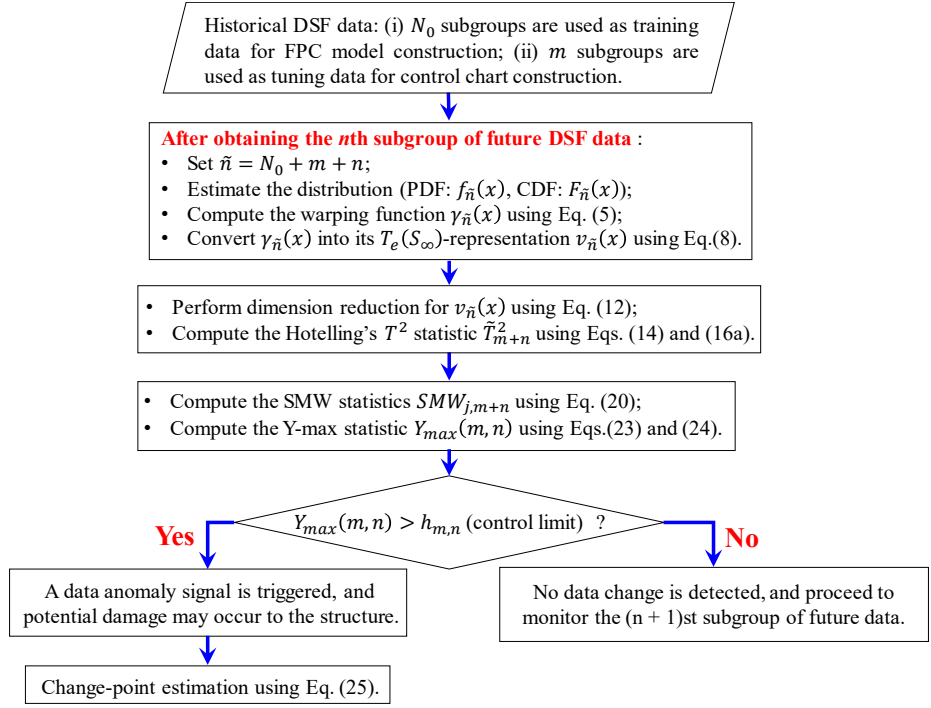

**Fig. 10.** The implementation procedure of the Warp-RANK-$T^2$ control chart for detecting distributional changes in DSF data.

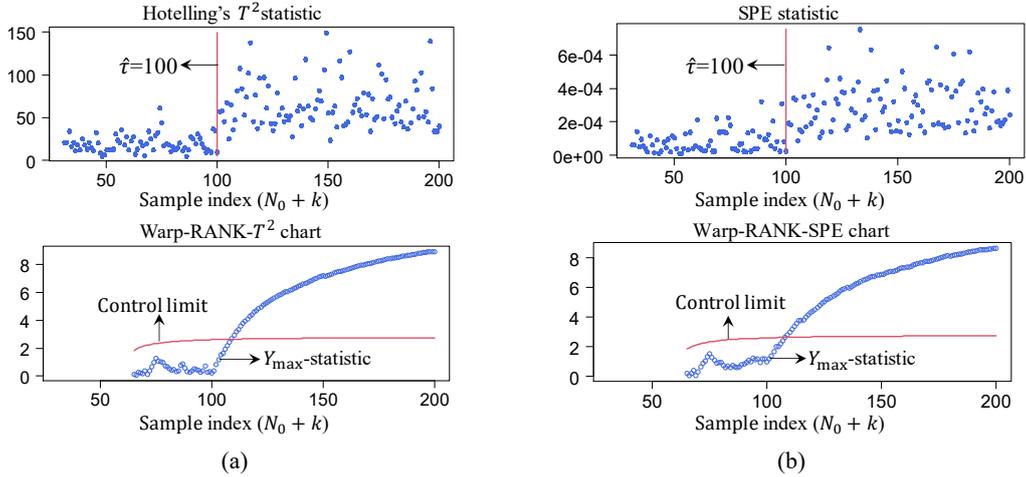

**Fig. 11.** The constructed control charts for the distributional data in Fig. 1. (a) Warp-RANK-$T^2$ chart and (b) Warp-RANK-SPE chart. The parameters for chart construction are: $N_0 = 30$ (training data length), $m_0 = 4$ (parameter in Eq. (23)), and $\lambda = 0.05$ (smoothing parameter in Eq. (23)). Control limits are derived from Table 1 of [54] with IC ARL=500. The estimated change point is marked by a vertical line.

To facilitate practitioners in implementing the proposed warping function-based control chart for detecting distributional changes in DSF data, Fig. 10 presents the step-by-step implementation procedure for the Warp-RANK-$T^2$ chart using a flowchart. The flowchart for the Warp-RANK-SPE chart is similar to that of the Warp-RANK-$T^2$ chart, requiring only the replacement of Hotelling's $T^2$ statistic with the SPE statistic. Thus, its flowchart is omitted to save space. For illustrative purposes, Fig. 11 displays the control charts constructed for the distributional data presented in Fig. 1 using the proposed method, with the estimated change point marked by a solid vertical line. Compared to the traditional control charts in Fig. 2, the proposed approach effectively detects and locates the change in the distributional sequence.

## 4. Simulation studies

### 4.1. Simulation study I

In this subsection, a simulation study is conducted to compare the detection power of the proposed method against competing approaches. In statistics, detection power refers to the probability that a statistical detector correctly identifies a true effect when it actually exists in the population. Specifically, for the problem of detecting distributional changes in DSF data, the detection power quantifies the probability that the detector accurately signals a change in the DSF distributional sequence when such a change genuinely occurs in the underlying process.

In the proposed method, the distributional change detection problem is reformulated as a warping function change detection problem. This is achieved by constructing control charts based on the FPCA results of $T_e(S_\infty)$-representations of warping functions. To demonstrate the advantages of this approach over directly using PDFs as monitoring characteristics, we implement a competing method that performs detection on PDF sequences rather than warping function sequences. In this alternative approach, FPCA is directly applied to PDFs, and control charts are constructed based on the FPCA results in a manner analogous to our warping function-based control chart method. Consequently, this competing method is referred to as the PDF-FPCA-CC method. Another method selected for comparison is the change-point detection (CPD) approach proposed by Lei et al.[42]. This method first embeds the PDFs into Bayes space and then detects distributional changes by applying a functional change-point detector to the resulting Bayes-space embeddings. This competing method is referred to as the Bayes-CPD method.

Before simulating data, essential terminology and notation are presented. The initial state prior to any distributional change is designated as the in-control (IC) state, while the subsequent state after such a change is termed the out-of-control (OC) state. The transition point between IC and OC states is called the change point. Let $\mathcal{F} = \{f_1, \cdots, f_\tau, \cdots, f_T\}$ represent the PDF sequence of size $T$ to be simulated, where the integer $\tau \in (1, T)$ serves as the change point. The PDFs before and after $\tau$ are designated as IC and OC PDF samples, respectively. Let $\text{Betapdf}(a, b)$ denote the PDF of the beta distribution with shape parameters $a$ and $b$, and let $\mathcal{U}(\alpha, \beta)$ represent the uniform distribution on the interval $[\alpha, \beta]$. The IC PDF samples (i.e. $\{f_i\}_{i=1}^{\tau}$) are independently generated using the following single beta distribution model:

$$f_i = \text{Betapdf}(a_i, b_i), \qquad i = 1, \cdots, \tau \qquad (27)$$

where $a_i \sim \mathcal{U}(10,14)$ and $b_i \sim \mathcal{U}(14,20)$. The OC PDF samples (i.e. $\{f_i\}_{i=\tau+1}^{T}$) are independently generated through the following mixing beta distribution model:

$$f_i = (1-\delta) \cdot \text{Betapdf}(a_i, b_i) + \delta \cdot \text{Betapdf}(c_i, d_i), \qquad i = \tau+1, \cdots, T \qquad (28)$$

where $a_i \sim \mathcal{U}(10,14), b_i \sim \mathcal{U}(14,20), c_i \sim \mathcal{U}(14,20), d_i \sim \mathcal{U}(20,25)$, and $\delta \in (0,1)$ represents the mixing coefficient. A comparison of the data-generating models in Eqs. (27) and (28) shows that the OC-state PDF model (Eq. (28)) is formed by superimposing a beta PDF onto the IC-state PDF model (Eq. (27)), with the mixing proportion being determined by the parameter $\delta$. As $\delta$ increases, the OC-state PDF deviates more significantly from the IC-state PDFs. Consequently, $\delta$ controls the strength of the distributional change. Based on the established data-generating models, two different scenarios are considered for synthetic PDF sequence generation:

Scenario I: Sequences of fixed length $T = 130$ with a change point at $\tau = 100$;

Scenario II: Sequences of fixed length $T = 200$ with a change point at $\tau = 100$.

Given a parameter $\delta \in (0,1)$ in Eq. (28), the data-generating process is repeated 100 times, producing 100 distinct distributional sequences per scenario. These sequences are used to compute empirical detection power at the specified $\delta$. Let $\mathcal{F}_k^\delta = \{f_{k,1}^\delta(x), \cdots, f_{k,i}^\delta(x), \cdots, f_{k,T}^\delta(x)\}$ denote the $k$th PDF sequence generated under Scenario I (or Scenario II) with parameter $\delta$. The complete set of 100 PDF sequences can be collectively represented as $\{\mathcal{F}_k^\delta\}_{k=1}^{100}$. After independently applying a change detection method (e.g. the proposed method or competing methods) to each sequence in $\{\mathcal{F}_k^\delta\}_{k=1}^{100}$, let $K_{Ch}^\delta$ denote the count of sequences where changes are successfully detected. The empirical detection power of the method at $\delta$ is then calculated as:

$$EDP(\delta) = \frac{K_{Ch}^{\delta}}{100}, \quad \delta \in (0,1) \tag{29}$$

The empirical detection power $EDP(\delta)$ serves as a quantitative metric for method performance: higher values indicate greater detection capability at the given $\delta$.

Next, the proposed method and the two competing methods are applied to detect changes in simulated distributional sequences under both Scenario I (130-length) and Scenario II (200-length). For the proposed method, the training sample size $N_0$ is set to 30. The first 30 PDFs following the training data are utilized as tuning samples (i.e. the tuning sample length $m = 30$). The parameter $m_0$ in Eq. (23) is set to 4, and the smoothing parameter $\lambda$ in Eq. (23) is set to 0.05. The control limits $h_{m,n}$ are obtained from Table 1 of [54] by setting IC ARL=500. The PDF-FPCA-CC method is essentially a variant of the proposed approach, with the key difference being that FPCA is applied to PDFs rather than $T_e(S_\infty)$-representations of warping functions. As a result, its computational procedure closely follows that of the proposed warping function-based method, and the parameter settings for FPCA and control chart implementation remain identical to those used in the warping function-based approach. For the Bayes-CPD method, the default settings from Lei et al. [42] are adopted, with the detector's significance level set to 0.1.

**Table 1** Empirical detection powers computed using the simulated dataset in Scenario I.

| | Empirical detection power $EDP(\delta)$ | | | | | | | | | | | |
|---|---|---|---|---|---|---|---|---|---|---|---|---|
| $\delta$ | 0.05 | 0.07 | 0.10 | 0.15 | 0.20 | 0.25 | 0.30 | 0.40 | 0.50 | 0.60 | 0.80 | 1.00 |
| Proposed method | 0.09 | 0.25 | 0.60 | 0.94 | 0.99 | 1.00 | 1.00 | 1.00 | 1.00 | 1.00 | 1.00 | 1.00 |
| PDF-FPCA-CC | 0.01 | 0.01 | 0.02 | 0.09 | 0.27 | 0.45 | 0.71 | 0.95 | 1.00 | 1.00 | 1.00 | 1.00 |
| Bayes-CPD | 0.04 | 0.04 | 0.04 | 0.04 | 0.04 | 0.04 | 0.06 | 0.14 | 0.18 | 0.24 | 0.53 | 0.97 |

**Table 2** Empirical detection powers computed using the simulated dataset in Scenario II.

| | Empirical detection power $EDP(\delta)$ | | | | | | | | | | | |
|---|---|---|---|---|---|---|---|---|---|---|---|---|
| $\delta$ | 0.05 | 0.07 | 0.10 | 0.15 | 0.20 | 0.25 | 0.30 | 0.40 | 0.50 | 0.60 | 0.80 | 1.00 |
| Proposed method | 0.33 | 0.83 | 0.99 | 1.00 | 1.00 | 1.00 | 1.00 | 1.00 | 1.00 | 1.00 | 1.00 | 1.00 |
| PDF-FPCA-CC | 0.08 | 0.11 | 0.16 | 0.31 | 0.78 | 0.95 | 1.00 | 1.00 | 1.00 | 1.00 | 1.00 | 1.00 |
| Bayes-CPD | 0.09 | 0.09 | 0.10 | 0.12 | 0.17 | 0.21 | 0.27 | 0.39 | 0.61 | 0.88 | 1.00 | 1.00 |

The computed empirical detection powers for Scenario I and Scenario II are presented in Table 1 and Table 2, respectively. Compared to the two competing methods, the proposed method demonstrates significantly higher detection power. The Bayes-CPD method exhibits reduced detection power when the post-change sample size is limited (Scenario I). In such cases, it requires a substantial distributional shift magnitude (quantified by $\delta$) to effectively identify data changes. While the performance of the method improves with larger post-change sample volumes (e.g. Scenario II with $n_{OC} = 100$), it still significantly underperforms compared to the proposed approach (Table 2). With greater detection power, the proposed approach's superior detection capability over competing methods is validated through the simulation results.

*4.2. Simulation study II*

In this subsection, an additional simulation study is conducted to demonstrate the robustness of the proposed method against outlying PDFs in the distributional sequence, while also comparing its performance with the direct charting method mentioned earlier in Subsection 3.4. The synthetic PDF sequence is generated using beta distribution models. The simulated PDF sequence has a fixed length of $T = 230$ with a change point at $\tau = 200$. Since the direct charting method requires relatively lengthy tuning data for control limit estimation, the simulated PDF sequence is longer than those in the previous simulation study. The IC PDF samples $\{f_i(x)\}_{i=1}^{\tau}$ (i.e. pre-change data) are independently generated by $f_i(x) = \text{Betapdf}(a_i, b_i)$, where $a_i \sim \mathcal{U}(10,14)$ and $b_i \sim \mathcal{U}(14,17)$. The OC PDF samples $\{f_i(x)\}_{i=\tau+1}^{T}$ (i.e. post-change data) are independently generated by another beta distribution model $f_i(x) = \text{Betapdf}(c_i, d_i)$, with $c_i \sim \mathcal{U}(14,18)$ and $d_i \sim \mathcal{U}(16,20)$. The outlying PDFs $\{f_k^{out}(x)\}_k^{N_{out}}$ (where $N_{out}$ denotes the number of outlying PDFs to be simulated) are generated using a third beta distribution model $f_k^{out}(x) = \text{Betapdf}(u_k, v_k)$, with $u_k \sim \mathcal{U}(12,16)$ and $v_k \sim \mathcal{U}(22,26)$. Let $\mathcal{F} = \{f_1(x), \cdots, f_\tau(x), \cdots, f_T(x)\}$ denote the simulated PDF sequence without outliers, where $\tau = 200$ marks the change point. Then, 4 PDFs in $\mathcal{F}$ indexed by $\{160, 161, 162, 163\}$ are replaced with outlying PDFs. These locally occurring anomalous PDFs are employed to simulate the short-range anomaly phenomena depicted in Fig. 3. As discussed in Section 2, after distribution summary, short-range anomalous data in the raw DSF data can be condensed into a few consecutive anomalous outlying PDFs within the distributional sequence.

The parameter settings for the proposed method maintain the same as those used in the previous simulation study. For the direct charting method, the charting statistics are the Hotelling's $T^2$ statistics $\{\tilde{T}_k^2\}$ (Eq. (16a)) and SPE statistics $\{\widetilde{SPE}_k\}$ (Eq. (16b)). The upper control limit (UCL) is estimated as the $1-\alpha$ sample quantile of the first 100 samples of $\{\widetilde{SPE}_k\}$ (or $\{\tilde{T}_k^2\}$), where $\alpha$ represents the overall Type I error probability. In this study, $\alpha$ is set to 0.01.

The control charts generated by the proposed approach and the direct charting method are presented in Fig. 12 and Fig. 13, respectively. As shown, the proposed method (Fig. 12) demonstrates robustness against outliers. The estimated change point from Eqs. (25) and (26) is $\hat{\tau} = 200$, which matches the true change point at $\tau = 200$. In contrast, the direct charting method (Fig. 13) is vulnerable to outliers, triggering false alarms at outlier locations before the true pattern change at $\tau = 200$.

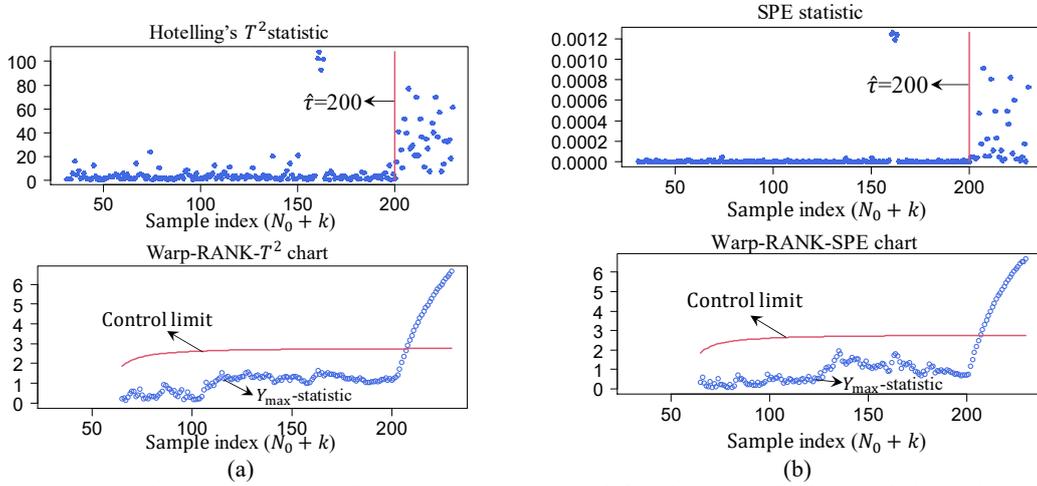

**Fig. 12.** Control charts constructed using the proposed method for simulated distributional data with outlier disturbances. (a) Warp-RANK-$T^2$ chart and (b) Warp-RANK-SPE chart. The estimated change point is marked by a vertical line.

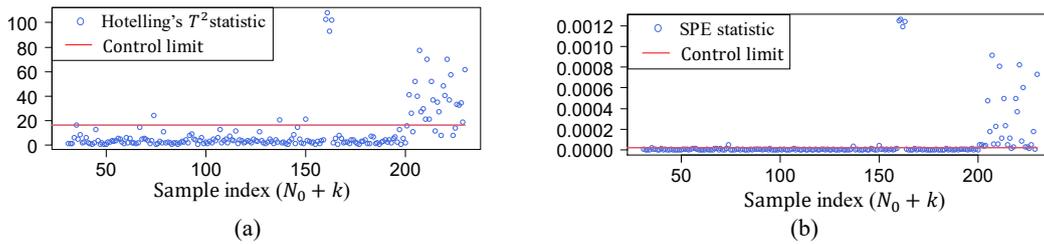

**Fig. 13.** Control charts generated by the direct charting method for simulated distributional data with outlier disturbances. (a) Control chart for the Hotelling's $T^2$ statistics and (b) control chart for the SPE statistics.

To investigate the impact of the smoothing parameter ($\lambda$ in Eq. (23)) on the warping function-based control chart, Fig. 14 presents Warp-RANK-$T^2$ charts constructed using six distinct smoothing parameter values: $\lambda = 0.10, 0.15, 0.20, 0.25, 0.30$, and $0.50$. As the smoothing parameter $\lambda$ increases, the robustness of the Warp-RANK-$T^2$ chart decreases. The results in Fig. 14 demonstrate that when $\lambda$ reaches a value as high as 0.30, false alarms may occur at locations disturbed by outliers. This occurs because a larger $\lambda$ value improves the influence of current data observations (represented by $SMW_{j,m+n}$) on the Y-max statistic computed via Eqs. (23) and (24), thereby increasing the chart's responsiveness to outliers. The smoothing parameter $\lambda$ exerts a similar effect on the Warp-RANK-SPE chart. Consequently, $\lambda$ can be regarded as a tuning parameter for adjusting the robustness of the proposed method. To enhance robustness, a smaller smoothing parameter value is recommended.

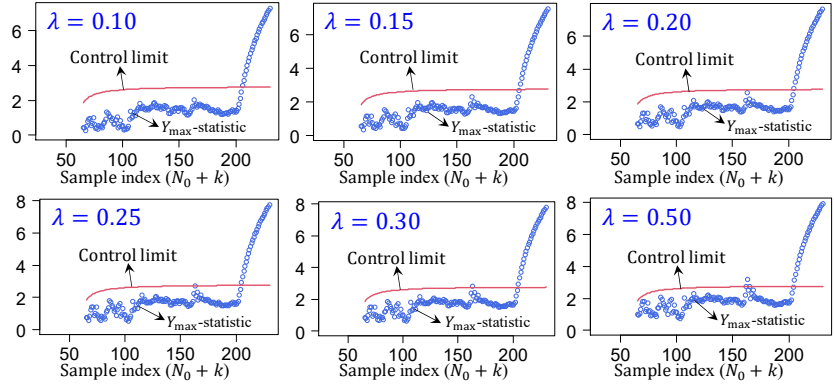

**Fig. 14.** Warp-RANK-$T^2$ charts constructed using different smoothing parameters ($\lambda = 0.10$, $\lambda = 0.15$, $\lambda = 0.20$, $\lambda = 0.25$, $\lambda = 0.30$ and $\lambda = 0.50$).

## 5. Case study based on SHM data of a real cable-stayed bridge

In this section, the proposed method is applied to structural condition assessment of a real long-span bridge to validate its effectiveness for real-world SHM applications. The monitoring data utilized in this study are tension measurements of stay cables collected by an SHM system installed on a cable-stayed bridge in China. This bridge has an arc-shaped steel tower, with a main span of 648 meters and side spans of 257 meters and 63 meters. The bridge deck accommodates six lanes of expressway traffic in both directions. The stay cable system of this bridge consists of 168 pairs of stay cables symmetrically installed on both sides (upstream and downstream) of the bridge deck, which transfer static and dynamic loads from the deck to the towers. To ensure operational safety, a large-scale SHM system was equipped to this bridge. For cable tension monitoring, each individual stay cable was equipped with a load cell. These sensors continuously collect tension data from the cables at a sampling frequency of 10 Hz.

Stay cables, as the core load-bearing components of cable-stayed bridges, directly affect the overall structural safety of the bridge. When stay cables sustain damage, it not only significantly reduces the bridge's load-bearing capacity but may also trigger cascading structural failure risks.

However, stay cables are usually encased in protective sheaths made of high-density polyethylene or steel. While this design effectively prevents environmental corrosion, it also creates a visual inspection barrier. Conventional visual inspection methods cannot penetrate these protective layers to detect internal potential damage. This 'invisibility' makes early damage identification of stay cables a technical challenge in bridge health monitoring, driving the need for non-contact inspection technologies to overcome this limitation. For this issue, data-driven structural damage identification or condition assessment approaches provide a compelling solution. By extracting DSFs from raw monitoring data, these methods transform the problem of damage detection or structural condition assessment into a problem of detecting changes in the distribution of the extracted DSF data. Advanced statistical analysis of monitoring data forms the foundation of these methodologies, as reliable statistical tools are essential for detecting subtle changes in DSF distributions. Such distributional changes often indicate damage, and their early detection is critical for issuing timely warnings before severe structural degradation occurs.

For cable-stayed bridges, the ratio of cable tensions between paired stay cables installed at identical cross-sections of the bridge deck demonstrates substantial potential as a DSF for cable condition assessment [56]. Existing studies demonstrate that the cable tension ratio (CTR) is insensitive to dynamic load variations on the bridge deck while remaining highly sensitive to changes in the structural condition of the stay cables [56]. As a result, alterations in the condition of the stay cables can be reflected in changes in the probability distribution of CTR data. For detailed discussion, refer to Li et al. [56]. This paper also adopts the CTR as the DSF. The collected CTR data are divided into segments, with probability distributions estimated for each segment to construct the CTR distributional sequence. Subsequently, the proposed warping function-based control charts are implemented to detect and monitor changes in the distribution of the CTR data.

Four pairs of stay cables (labeled SCP-a, SCP-b, SCP-c, and SCP-d, where 'SCP' stands for 'stay-cable pair') are selected for investigation, as indicated by the bold solid lines in Fig. 15. For each cable pair, 170 days of tension monitoring data are extracted for subsequent analysis. For CTR computation, synchronized data from both sensors on the same cable pair must be used. Consequently, only data with complete records from both sensors of a cable pair are selected. Additionally, due to sensor malfunctions or external disturbances, raw monitoring data contain invalid records or gaps during certain time periods. Such data are excluded from further analysis.

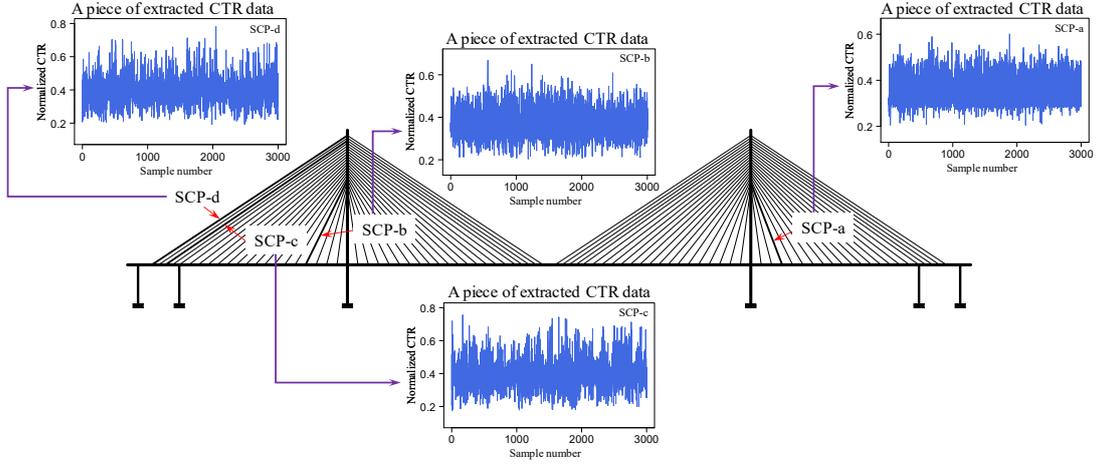

**Fig. 15.** Selected cable pairs (i.e. SCP-a, SCP-b, SCP-c, and SCP-d, indicated by bold lines) for investigation, along with their corresponding representative extracted CTR samples (after preprocessing and scaling to the unit interval [0,1]; only one piece of extracted CTR data is presented for each cable pair).

The calculation of CTR data follows the same procedure as in Li et al. [56], with specific implementation details omitted for brevity. The CTR data are presented in their non-logarithmic form, similar to that in Lei et al. [42]. It is important to note that, prior to CTR computation, raw monitoring data must be preprocessed using the method described in Section 3 of Li et al. [56] to eliminate ambient variations. This preprocessing step constitutes the first stage of the statistical pattern recognition paradigm for damage detection, namely data normalization. The subsequent CTR calculation represents the second stage of this paradigm, namely DSF extraction. From the extracted 170-day CTR dataset, the first 30-day period is designated as training data for estimating the support of CTR distributions and constructing the FPC model. Before estimating the support, a boxplot-based outlier detection method is applied to the training data to identify and exclude extreme outliers (i.e. those with extremely low or high values). This preprocessing step is critical because extreme outliers can cause the estimated support to be substantially broader than the true support. Let $CTR_{tr} = \{Z_1, Z_2, \cdots, Z_{N_{tr}}\}$ denote the 30-day training CTR data (where $N_{tr}$ is the sample size) after extreme outlier removal for a given cable pair. Let $s_{N_{tr}}$ be the sample standard deviation of $CTR_{tr}$. The lower and upper bounds of the training data population are estimated as $\widehat{LB} = \min_{1 \leq i \leq N_{tr}} Z_i - s_{N_{tr}}/\sqrt{N_{tr}}$ and $\widehat{UB} = \max_{1 \leq i \leq N_{tr}} Z_i + s_{N_{tr}}/\sqrt{N_{tr}}$, respectively [48]. The support of the CTR distribution is initially estimated by the interval $[\widehat{LB}, \widehat{UB}]$. However, this interval is not ideal because future data may experience mean shifts or variance changes, rendering the training data-based estimate potentially inapplicable. To address this, the interval $[\widehat{LB}, \widehat{UB}]$ is appropriately expanded to produce a broader support estimate, thereby reducing the risk of failing to cover future data variations. Consequently, the estimated support of the CTR distribution is constructed as

$$[\widehat{LB}^*, \widehat{UB}^*] = [\widehat{LB} - \theta(\widehat{UB} - \widehat{LB}), \widehat{UB} + \theta(\widehat{UB} - \widehat{LB})] \tag{30}$$

where $\theta$ is a tuning parameter that controls the degree of interval widening, with a default value of 0.4. Given the estimated support interval $[\widehat{LB}^*, \widehat{UB}^*]$, the CTR data $\{Z_1, Z_2, \cdots\}$ are further scaled to the unit interval [0,1] as follows:

$$\tilde{Z}_i = \frac{Z_i - \widehat{LB}^*}{\widehat{UB}^* - \widehat{LB}^*}, \quad i = 1, 2, \cdots \tag{31}$$

Fig. 15 shows a subset of these normalized CTR data for each selected cable pair.

The scaled data from Eq. (31) are divided into daily segments, and the PDFs for each day are estimated using kernel density estimation. To ensure the estimated PDFs are supported on the finite interval [0,1], the modified kernel density estimator proposed in [57] is used. The Gaussian kernel is selected, and the bandwidth parameter is adaptively determined using Silverman's rule of thumb [58]. This process corresponds to the data subgrouping and distribution summary procedure described in Section 2, resulting in a sequence of 170 PDFs per cable pair. The distributional sequences are denoted as $\{f_i^a(x)\}_{i=1}^{170}$, $\{f_i^b(x)\}_{i=1}^{170}$, $\{f_i^c(x)\}_{i=1}^{170}$ and $\{f_i^d(x)\}_{i=1}^{170}$ for cable pairs SCP-a, SCP-b, SCP-c and SCP-d, respectively. The first 30 PDFs in each sequence, corresponding to the 30-day training data, are designated as training PDFs for constructing the FPC model, corresponding to $N_0 = 30$ in Eq. (9). As mentioned in Subsection 3.3, these training data are assumed to precede structural changes. Thus, the focus is primarily on detecting and monitoring changes in the remaining data. For clarity, the remaining 140 PDFs in each distributional sequence (after excluding the 30 training PDFs) are re-expressed as $\{\tilde{f}_k^a(x)\}_{k=1}^{140}$, $\{\tilde{f}_k^b(x)\}_{k=1}^{140}$, $\{\tilde{f}_k^c(x)\}_{k=1}^{140}$ and $\{\tilde{f}_k^d(x)\}_{k=1}^{140}$, which are referred to as the test (distributional) data. The curves of the test PDFs for the four cable pairs are visualized in the left column of Fig. 16, while the right column of Fig. 16 presents the corresponding heatmap plots of the distributional sequences.

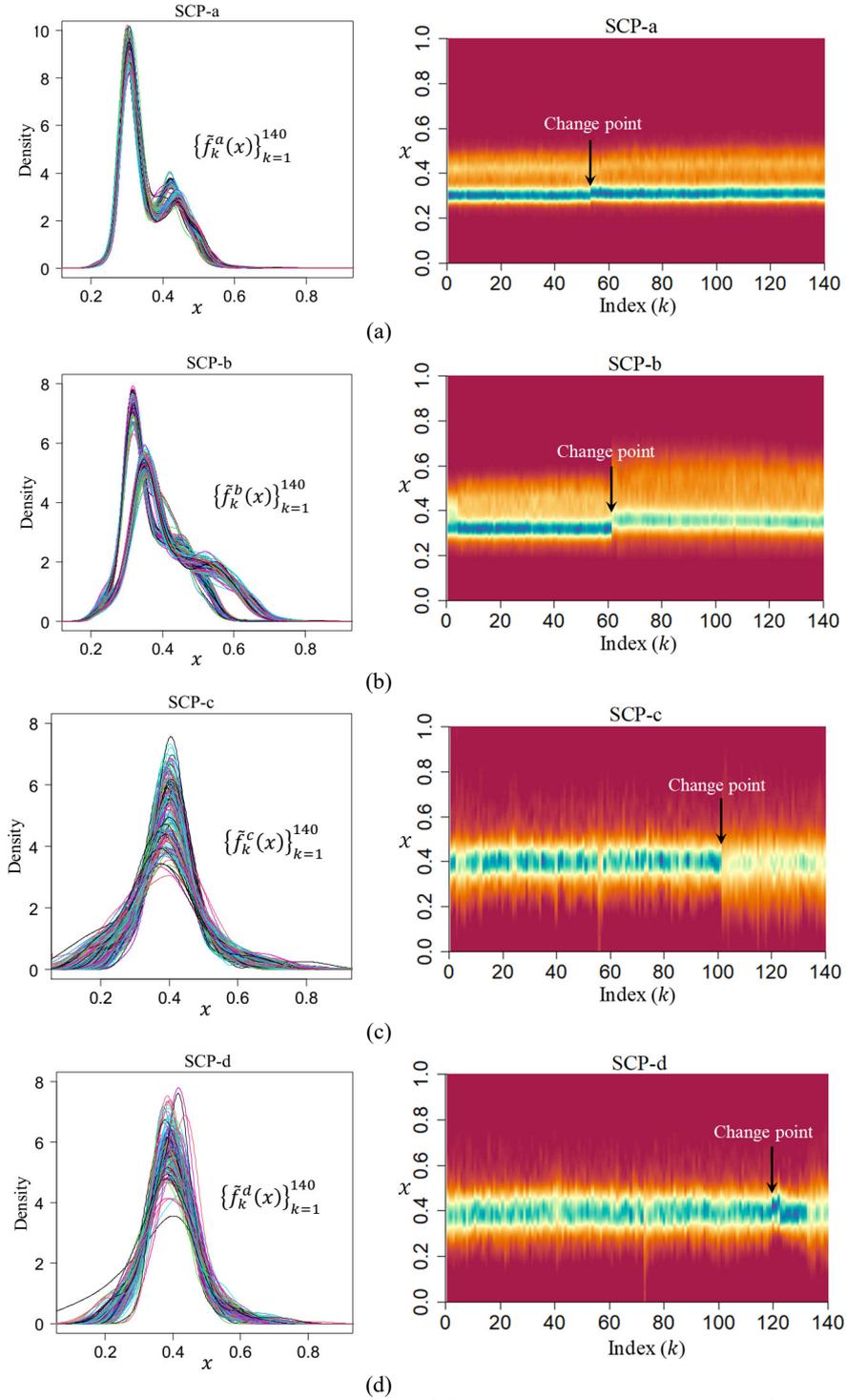

**Fig. 16.** Visualizations of distributional sequences: (a) $\{\tilde{f}_k^a(x)\}_{k=1}^{140}$ (cable pair SCP-a), (b) $\{\tilde{f}_k^b(x)\}_{k=1}^{140}$ (cable pair SCP-b), (c) $\{\tilde{f}_k^c(x)\}_{k=1}^{140}$ (cable pair SCP-c) and (d) $\{\tilde{f}_k^d(x)\}_{k=1}^{140}$ (cable pair SCP-d). The left column corresponds to the curves of the PDFs, while the right column corresponds to the heatmap representations of PDF sequences.

As depicted in the heatmap representations of distributional sequences (right column of Fig. 16), all four extracted sequences exhibit distinct pattern changes. The sample indices corresponding to these pattern changes are referred to as change points and are individually marked with arrows in each plot. Notably, the distributional sequences corresponding to cable pairs SCP-b and SCP-c demonstrate particularly significant changes, which can be considered as strong change cases. In contrast, the pattern changes in the distributional sequences corresponding to cable pairs SCP-a and SCP-d are relatively subtle, which can be considered as weak change cases. In the following, the proposed method will be applied to detect these distributional changes and validate its effectiveness in practical engineering applications.

Similar to the simulation study, after extracting warping functions from the distributional data and mapping them into the tangent space, the resulting functional data undergoes FPC decomposition. For each distributional sequence, the FPC model is established using the training data. Both the Hotelling's $T^2$ and SPE statistics constructed based on the FPC results are considered as monitoring features. The Y-max statistic, derived from the combination of Eqs. (23) and (24), functions as the charting statistic and can also serve as a damage index for damage detection or structural condition assessment. When the Y-max statistic exceeds the control limit, an alarm is triggered, and the change point (i.e. the location of the distributional change) is estimated using Eq. (25). For the four distributional sequences, the parameter settings for warping function extraction, FPCA, and control chart construction remain identical to those in Simulation Study I.

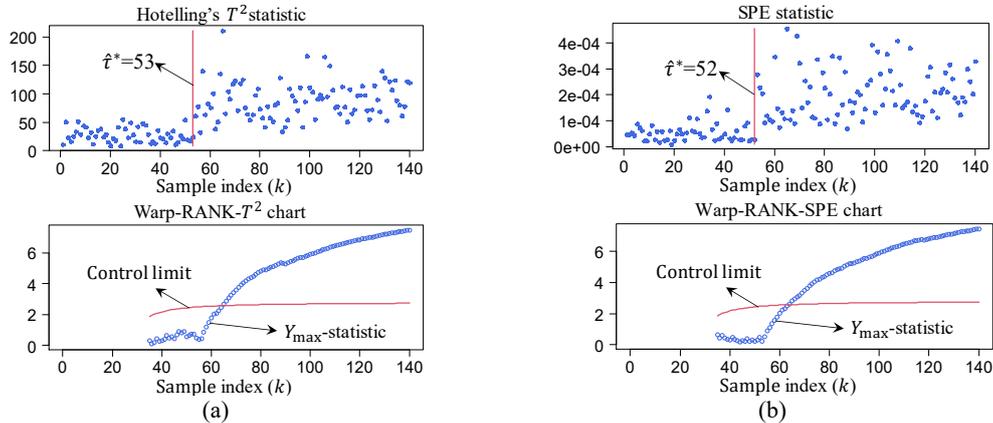

**Fig. 17.** Control charts for the distributional data of cable pair SCP-a. (a) Warp-RANK-$T^2$ chart and (b) Warp-RANK-SPE chart. The estimated change point is marked by a vertical line.

The constructed control charts and change-point estimation results for cable pairs SCP-a, SCP-b, SCP-c, and SCP-d are presented in Fig. 17, Fig. 18, Fig. 19 and Fig. 20, respectively. The estimated change-point locations are marked with vertical lines. Through manual examination, the true change points of the distribution sequences for cable pairs SCP-a, SCP-b, SCP-c, and SCP-d are found at $T = $ 53, 61, 101, and 119, respectively. The change-point estimates closely match the true change points, indicating that the proposed method delivers reliable location estimates for all four change points. Notably, even in weak-change scenarios (e.g. SCP-a and SCP-d), the method not only issues timely warnings but also achieves high-precision change-point estimates.

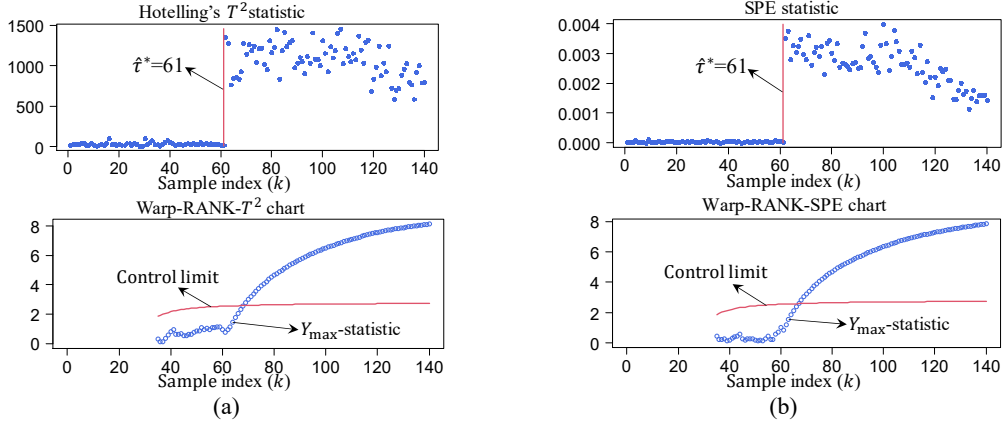

**Fig. 18.** Same as Fig. 17, except that the control charts are constructed for the data from cable pair SCP-b.

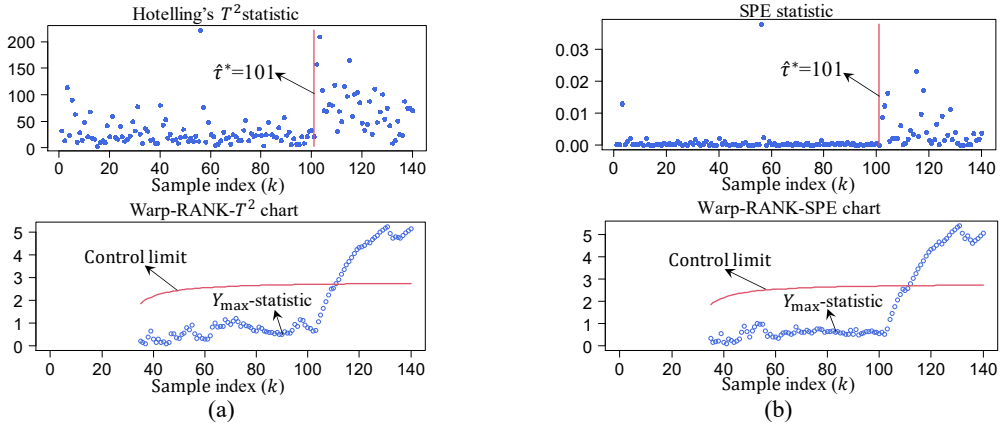

**Fig. 19.** Same as Fig. 17, except that the control charts are constructed for the data from cable pair SCP-c.

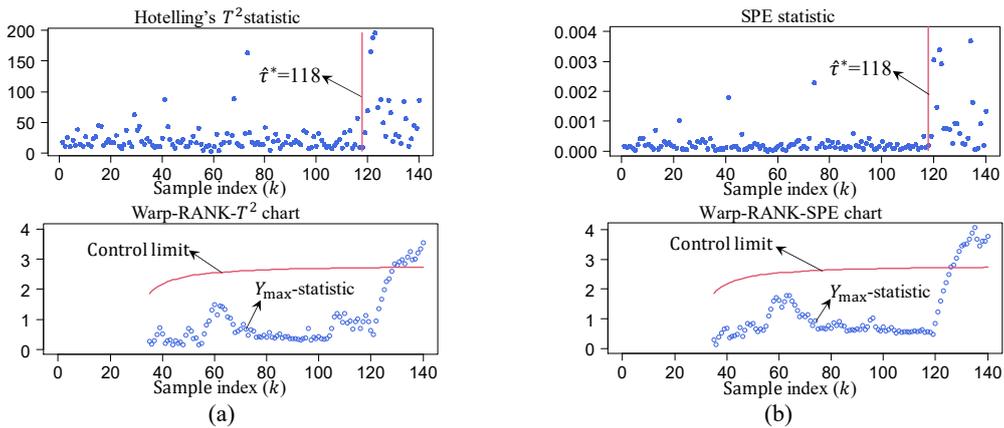

**Fig. 20.** Same as Fig. 17, except that the control charts are constructed for the data from cable pair SCP-b.

Finally, the identified change point is used to divide each distributional sequence into pre-change and post-change segments. Fig. 21 presents the $T_e(S_\infty)$-representations of the corresponding warping functions, clearly demonstrating discriminability between pre- and post-change data. For the case of weak changes shown in Fig. 16 (a) and (d), the changes are not easily noticeable by visually examining the original distributional sequences. However, the proposed method enhances detectability in Hotelling's $T^2$ or SPE statistic sequences derived from the $T_e(S_\infty)$-representations of warping functions, making subtle changes more distinguishable. This further validates the effectiveness of the proposed method.

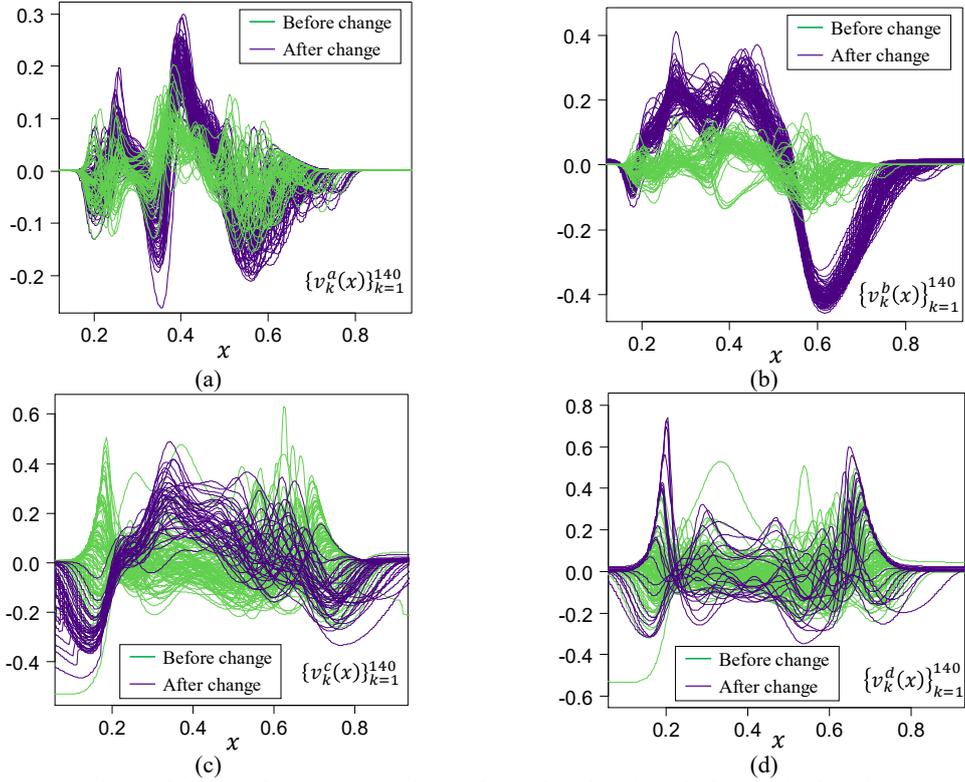

**Fig. 21.** Comparison of the $T_e(S_\infty)$-representations of warping functions before and after the pattern change (demarcated by the estimated change point). (a) Data from cable pair SCP-a, (b) Data from cable pair SCP-b, (c) Data from cable pair SCP-c, and (d) Data from cable pair SCP-d. The data before the change point is represented by light curves, while the data after the change point is represented by dark curves.

As illustrated in Fig. 17, Fig. 19 and Fig. 20, the extracted Hotelling's $T^2$ or SPE statistic sequences contain randomly distributed outliers (particularly evident in Fig. 20). These outliers correspond to anomalous observations in the real datasets. The sporadic nature of these anomalies suggests they are primarily caused by random disturbances rather than indicative of structural damage. In real-world SHM applications, such anomalous samples are very common and often unavoidable. These samples can interfere with detection results and trigger false alarms. However, the proposed method demonstrates strong robustness against such anomalies. Crucially, the corresponding alarms are only triggered when the actual pattern change occurs in the distributional sequence, rather than being immediately activated by the presence of these sporadic anomalous samples.

Since the CTR data serve as DSFs of stay cables, the changes detected by the control chart in the CTR distributional sequences may indicate structural condition alterations, such as damage or performance degradation. In practical applications, when an anomaly alarm is triggered by the control chart and the occurrence time of the change is pinpointed by the estimated change point, SHM practitioners should conduct a thorough investigation into the underlying causes. The investigation aims to clarify whether these distributional changes stem from structural damage or other sudden disturbances (e.g. operational or environmental factors). Generally, if no evidence supports the latter explanation, the detected change is most likely induced by damage.

## 6. Conclusions

This paper proposes a novel control chart designed to detect distributional changes in damage-sensitive feature (DSF) data, which is used for data-driven structural condition assessment. The proposed control chart is constructed using warping transformation of distributions, functional principal component analysis (FPCA), and rank-based nonparametric statistical methods. This approach enables the detection of complex distributional changes and demonstrates satisfactory robustness. The practical performance of the proposed method is evaluated through both simulation and real-data studies. The following conclusions are drawn:

(1) The proposed method can not only detect shifts in the probability distribution of DSF data but also identify its complex shape changes, thereby addressing the limitations of traditional control charts in detecting complex distributional changes of DSF data. This capability primarily stems from the innovative application of warping functions to capture distributional deformations and shifts, allowing the extraction and analysis of complex changes in distribution through warping functions. Online detection and warnings for complex distributional feature changes can be achieved by applying the proposed warping function-based control chart. Simulation studies indicate that this method possesses superior detection power compared to direct detection of the PDFs.

(2) The proposed method combines the functionalities of control charts and change point detection. Like control charts, it continuously monitors the distributional characteristics of DSF data and triggers alerts when anomalies are detected. Additionally, it can locate the timing of distributional changes (i.e. change-point estimation). This enables a more efficient response to sudden changes in the distribution of DSF data, making the method particularly well-suited for anomaly detection in a stream of probability distributions derived from SHM data.

(3) The proposed method demonstrates strong resistance to data anomalies caused by external interference, effectively reducing the risk of false alarms in damage detection. This is mainly due to the following mechanisms: (a) The distribution summary strategy helps mitigate the impact of outlying measurements; (b) The rank-based charting statistic improves detection robustness; (c) The anti-interference capability of the control chart for warping function monitoring can be flexibly adjusted by tuning its smoothing parameter. Typically, a smaller smoothing parameter value enhances anti-interference capability, while a larger value weakens it.

## Appendix 1. Synthetic data-generation process for results in Figs. 1-2

The data for the results in Figs. 1-2 are generated using Algorithm 1.

---
**Algorithm 1**: Synthetic Data Generation for Results in Figs. 1-2
---
1: Independently generate 25,000 random samples from the beta distribution $\text{Beta}(5, 7)$ and denote the result as $X = \{X_1, \cdots, X_{25000}\}$.
2: Independently generate 25,000 random samples from the mixture beta distribution $0.7 \times \text{Beta}(5, 7) + 0.3 \times \text{Beta}(25, 9)$ and denote the result as $Y = \{Y_1, \cdots, Y_{25000}\}$.
3: Set $Z = \{(X - m_X)/s_X, (Y - m_Y)/s_Y\}$ to form a time series, where $m_X$ and $m_Y$ are the sample means of $X$ and $Y$, respectively, and $s_X$ and $s_Y$ are the sample standard deviations of $X$ and $Y$, respectively.
4: Divide $Z$ into 200 subgroups of equal length.
---